\documentclass[%
 aip,
 amsmath,amssymb,
 reprint,
]{revtex4-1}

\usepackage{graphicx}
\usepackage{dcolumn}
\usepackage{bm}
\usepackage{xcolor}
\usepackage[utf8]{inputenc}
\usepackage[T1]{fontenc}
\usepackage{mathptmx}
\usepackage{etoolbox}
\usepackage{tikz}
\usetikzlibrary{tikzmark}
\usepackage[utf8]{inputenc}
\usepackage[T1]{fontenc}
\usepackage{mathptmx}
\usepackage{etoolbox}
\usepackage{booktabs}  
\usepackage{caption} 
\usepackage{amsmath}
\usepackage{amssymb}

\usepackage{tikz}
\usetikzlibrary{shapes.geometric} 
\usetikzlibrary{arrows.meta}      
\usetikzlibrary{positioning}       
\usetikzlibrary{decorations.pathreplacing} 
\usetikzlibrary{backgrounds}    

\begin{document}


\title{Flat interface between amorphous ices and the role of MDA-like intermediate states in the LDA--HDA transformation}

\author{Anastasiia Shupletsova}
\email[]{garkul.aa@phystech.edu}
\affiliation{Joint Institute for High Temperatures of RAS, 125412 Moscow, Russia}

\author{Vladimir Stegailov}
\email[]{stegailov@jiht.ru}
\affiliation{Joint Institute for High Temperatures of RAS, 125412 Moscow, Russia}


\begin{abstract}
The pressure-induced transformation between low-density amorphous ice (LDA) and high-density amorphous ice (HDA) is a prototypical polyamorphic transition, yet its microscopic mechanism -- and in particular the nature of intermediate amorphous (IA) states -- has long remained unresolved.
We present the first characterisation of a flat LDA$~||~$HDA interface, determining its equilibrium thickness and showing its link to medium-density amorphous ice (MDA). Interestingly, the thickness of the interfacial layer remains constant under compression, while its position shifts reversibly into the LDA region -- an elastic response that exhibits kinetic hysteresis upon decompression, reminiscent of memory effects.
For high-precision discrimination of amorphous ices, we use a neural network classifier based on SOAP (Smooth Overlap of Atomic Positions) descriptors. The systematic description of the local structure emphasizes the importance of orientational information from the hydrogen bond network.
Furthermore, this approach enables molecular dynamics analysis of the LDA-to-HDA transformation, directly mapping the spatial distribution of intermediate structures relative to growing HDA domains and the surrounding LDA matrix, and revealing that MDA-like IA configurations are not a distinct bulk phase but rather localise at the LDA--HDA interface. 

\end{abstract}

\maketitle

\section{\label{sec:intro}Introduction}
In scientific and technological research, the study of the properties and structure of water in various states plays a fundamental role. Identifying the local structure of aqueous systems is challenging given the wide variety of different phases of water and its intricate phase diagram. Nevertheless, this is a very important issue for understanding the mechanisms of phase, polymorphic and polyamorphous transformations at the microscopic level~\cite{tanaka2019revealing}. Transformations between low- and high-density amorphous ice have been extensively discussed in the literature over the past 40 years~\cite{mishima1985apparently, gromnitskaya2001ultrasonic, brazhkin2003high, martovnak2005evolution, gallo2016water}, but still remain a subject of debate~\cite{tanaka2020liquid, amann2023liquid, gartner2022liquid,eltareb2024continuum, dhabal2024liquid, mokshin2024liquid, gallagher2026local}. 
However, despite the long history of studying amorphous-amorphous transitions, the vast majority of studies, both experimental and theoretical, have focused either on the bulk properties of individual phases or on the macroscopic kinetics of the transformation as a whole. The structure of the interface between coexisting amorphous phases, as well as its role in the transition mechanism, has remained virtually outside the field of view of researchers. This contrasts with the situation in other materials, where interface analysis is a standard tool for identifying the nature of a phase transition~\cite{allen1987cascade, fryer2001dependence, napolitano2017glass, brandl2013structure, galenko2005diffuse, di2008interplay, reichenbach2021solid}. 
The lack of similar studies for amorphous ices is due to both the experimental difficulties of directly observing nanoscale inhomogeneities and the complexity of modeling an equilibrium or metastable interface between two disordered condensed phases. At the same time, important precedents have accumulated for other amorphous materials demonstrating the fruitfulness of this approach~\cite{moras2018shear, reichenbach2021solid}. 

The landscape of amorphous ice physics was fundamentally altered by the recent discovery of medium-density amorphous ice (MDA)~\cite{rosu2023medium}. This discovery was significant not only for identifying a new amorphous ice form, but also for suggesting that the LDA -- HDA relationship might involve more than two distinct states, motivating scientists to further dive into the details of amorphous-amorphous transformations. There is an opinion that MDA is a nanocrystalline material (not glass like LDA and HDA) and has nothing to do with polyamorphism~\cite{tonauer2023nucleation}. Although recent computational studies by De Almeida Ribeiro et al. have demonstrated that MDA-like structures can be produced by shear deformation of Ih, LDA, or HDA ice at constant shear rates~\cite{de2024medium}. Their results suggest that MDA is a shear-induced state of amorphous ice (SDA) that occupies a continuous density range between the LDA and HDA limits, with the specific density determined by the applied shear rate. This continuum concept is also supported by Eltareb~et~al.~\cite{eltareb2024continuum}. 
Machine-learning-based structural analysis has provided additional insights into the nature of MDA. Faure Beaulieu~et~al. employed rotationally-invariant, high-dimensional order parameters and neural network classifiers to characterize local environments in LDA, HDA, and MDA~\cite{faure2024high}. Their results highlight the difficulty in unambiguously differentiating MDA from LDA, suggesting significant structural similarities between these forms. This structural ambiguity supports the interpretation of MDA as occupying an intermediate position in a continuum of amorphous structures rather than representing a sharply distinct phase.

A key methodological issue in the analysis of amorphous systems is the selection of an adequate description of the local atomic environment. Unlike crystalline materials, where the structure can be characterized by a unit cell and symmetry group, amorphous phases lack long-range order, and their identification is only possible through statistical distributions of local configurations~\cite{debenedetti2001supercooled, berthier2011theoretical}. Therefore, a compact, yet informative method for representing the local environment of each molecule is needed, one that is sensitive to structural differences between phases and simultaneously invariant with respect to translations, rotations, and atomic permutations.

Several approaches to solving this problem have been proposed in the literature. Traditional low-dimensional order parameters, such as tetrahedrality~\cite{errington2001relationship, chau1998new} and local structural indices~\cite{errington2001relationship}, have been successfully used to characterize the local structure. However, their accuracy and completeness prove to be insufficient to reveal the subtle features of the local atomic environment characteristic of structurally related amorphous phases, such as LDA and HDA.

High-dimensional atom-centered descriptors combined with machine learning offer a powerful alternative. This family includes bond-orientational order (BOO) parameters~\cite{steinhardt1983bond, zeng2022bond}, atom-centered symmetry functions (ACSF)~\cite{behler2007generalized, behler2011atom, behler2021four, gastegger2018wacsf}, and the Smooth Overlap of Atomic Positions (SOAP) descriptor~\cite{bartok2013representing}. Although BOO has been used successfully to classify LDA and HDA~\cite{martelli2020connection, faure2024high}, it only captures angular information. ACSF, though powerful, requires extensive empirical parameter tuning. SOAP overcomes these limitations by expanding the full local atomic density in both radial and angular bases, providing a complete and systematically convergent representation.
A related line of research has focused on comparing different structural indicators for supercooled water. Foffi and Sciortino demonstrated that many commonly used indicators become strongly correlated with density fluctuations near the liquid-liquid critical point, and some even display identical distributions~\cite{foffi2022correlated}. This finding underscores the need for a descriptor that does not rely on a small set of pre-selected order parameters -- a requirement naturally met by the SOAP approach, which encodes the full local atomic density.

For hydrogen-bonded systems, such as amorphous ices, a key advantage of SOAP is its ability to include both oxygen and hydrogen atoms in the description. This allows the descriptor to distinguish not only density characteristics but also subtle features of the hydrogen network, including variations in the number and orientation of hydrogen bonds. As shown in recent studies, taking the hydrogen environment into account is crucial for discriminating between LDA and HDA~\cite{gallagher2026local}.

However, the task of developing a descriptor for classifying local structure in the context of the continuum of intermediate amorphous ices is both extremely important and incredibly complex, bordering on the philosophical question of classifying continuous phenomena. Different variations of IA can be structurally very close to both LDA and HDA. Thus, it becomes more practical to develop a universal coordinate system (set of descriptors) for mapping the local structure of any amorphous ice and determining its ``coordinates'' in the broad amorphous landscape between LDA and HDA rather than searching for a ``descriptor for IA''. Such coordinates for structure identification could be, for example, the probabilities of belonging to the LDA or HDA phase~\cite{gallagher2026local, faure2024high}. 

Machine learning methods have shown significant progress in tasks that require efficient recognition and classification of various structures~\cite{schmidt2019recent, reiser2022graph, kapsiani2021random, takahashi2021searching, doi2019machine, doi2021searching, boattini2019unsupervised, gallo2022unsupervised, ishiai2023graph, ishiai2024novel, donkor2024beyond}. However, the range of different ML techniques and approaches makes one wonder which one is most optimal. The most popular approach is the use of neural networks~\cite{martelli2020connection, monserrat2020liquid, geiger2013neural, faure2024high, ramesh2024microscopic} that can be trained to detect the local arrangement of atoms by different descriptors. 

In this work, we employ a neural network classifier trained on SOAP descriptors to analyze the local structure of amorphous ices during the pressure-induced LDA~$\to$~HDA transformation. We focus on two key aspects: (i) the evolution of the phase composition during compression, with particular attention to the role of intermediate states, and (ii) the structure and properties of the LDA~||~HDA flat interface, including its thickness, composition, and response to external pressure variations. Our approach allows us to quantitatively characterize the interface and to demonstrate that IA states are localized at the phase boundary rather than forming a bulk intermediate phase.

\section{\label{sec:method}Methods}
\begin{figure*}[ht]
\centering
\begin{tikzpicture}[
    arrow/.style={->, thick, >=Stealth},
    vector/.style={draw, thick, rounded corners, inner sep=8pt, fill=green!10},
    neuron/.style={circle, draw, minimum size=0.35cm, inner sep=0pt, fill=blue!30},
    labelnode/.style={align=center, font=\tiny}
]

\node (img) at (0,0) {\includegraphics[trim=55 75 48 75, clip, width=0.45\textwidth]{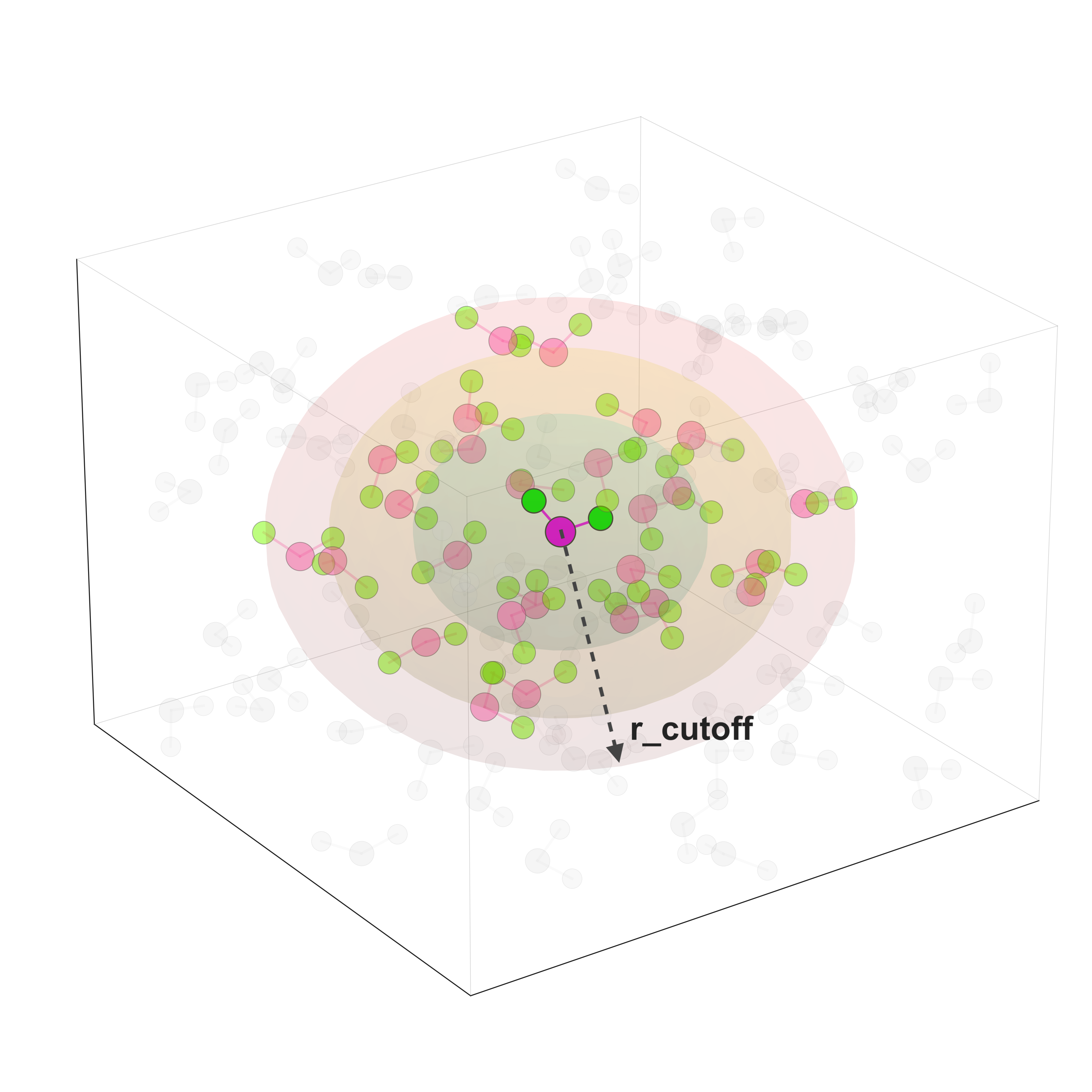}};

\draw[arrow] (2.0,0) -- (3.3,0);
\node[below] at (2.6,-0.05) {SOAP};

\node[vector] (vec) at (4.8,0) {
    $\mathbf{f}(\mathbf{r}_i) = 
    \begin{pmatrix}
    f^{\text{OO}}_{n_1 n_2 l} \\[4pt]
    f^{\text{HH}}_{n_1 n_2 l} \\[4pt]
    f^{\text{OH}}_{n_1 n_2 l}
    \end{pmatrix}$
};

\node[below=4pt of vec.south, align=center] {
    $0 \leq n_1 \leq n_2 \leq n_{\text{max}}$ \\
    $0 \leq l \leq l_{\text{max}}$
};

\begin{scope}[xshift=6.1cm]

\foreach \i in {-0.8,0,0.8} {
    \foreach \j in {-1.2,-0.4,0.4,1.2} {
        \draw[very thin, gray!40] (0, \i) -- (1.1, \j);
    }
}

\foreach \i in {-1.2,-0.4,0.4,1.2} {
    \foreach \j in {-1.2,-0.4,0.4,1.2} {
        \draw[very thin, gray!40] (1.1, \i) -- (2.2, \j);
    }
}

\foreach \i in {-1.2,-0.4,0.4,1.2} {
    \foreach \j in {-0.8,0,0.8} {
        \draw[very thin, gray!40] (2.2, \i) -- (3.3, \j);
    }
}

\foreach \i in {-0.8,0,0.8} {
    \node[neuron] (input-\i) at (0, \i) {};
}
\node[labelnode, font=\scriptsize, rotate=90] at (0, 1.5) {Input};

\foreach \i in {-1.2,-0.4,0.4,1.2} {
    \node[neuron, fill=purple!30] (hidden1-\i) at (1.1, \i) {};
}
\node[labelnode, font=\scriptsize, rotate=90] at (1.1, 2.0) {Hidden 1};

\foreach \i in {-1.2,-0.4,0.4,1.2} {
    \node[neuron, fill=purple!30] (hidden2-\i) at (2.2, \i) {};
}
\node[labelnode, font=\scriptsize, rotate=90] at (2.2, 2.0) {Hidden 2};

\foreach \i in {-0.8,0,0.8} {
    \node[neuron, fill=orange!40] (output-\i) at (3.3, \i) {};
}
\node[labelnode, font=\scriptsize, rotate=90] at (3.3, 1.5) {Output};

\draw[arrow] (3.6,0) -- (4.7,0);
\node[below, font=\small] at (4.1, -0.05) {Softmax};

\begin{scope}[shift={(6.95, -1.2)}, scale=1.5]
    \draw[fill=orange!50] (-0.9,0) rectangle (-0.5,1.4);
    \draw[fill=green!50] (-0.4,0) rectangle (0.0,0.5);
    \draw[fill=blue!50] (0.1,0) rectangle (0.5,0.9);
    
    \draw[->] (-1.3,0) -- (1.0,0);
    \draw[->] (-1.1,0) -- (-1.1,1.7);
    
    \node[font=\scriptsize, rotate=90] at (-0.7,0.7) {LDA};
    \node[font=\scriptsize, rotate=90] at (-0.2,0.25) {MDA};
    \node[font=\scriptsize, rotate=90] at (0.3,0.45) {HDA};
    
    \node[left, font=\small, rotate=90] at (-1.25,1.33) {probability};
\end{scope}

\end{scope}

\end{tikzpicture}
\caption{SOAP descriptor construction (left) followed by the neural network classification pipeline (right). Each water molecule is represented by its Cartesian coordinates $(x, y, z)$. The SOAP transformation maps the local atomic environment into a high-dimensional feature vector $\mathbf{f}(\mathbf{r}_i)$ comprising OO, HH, and OH correlations with radial indices $n_1, n_2$ and angular index $l$. The neural network has two hidden layers (600 neurons each). The output layer produces logits that are converted via softmax to class probabilities.}
\label{fig:combined_pipeline}
\end{figure*}
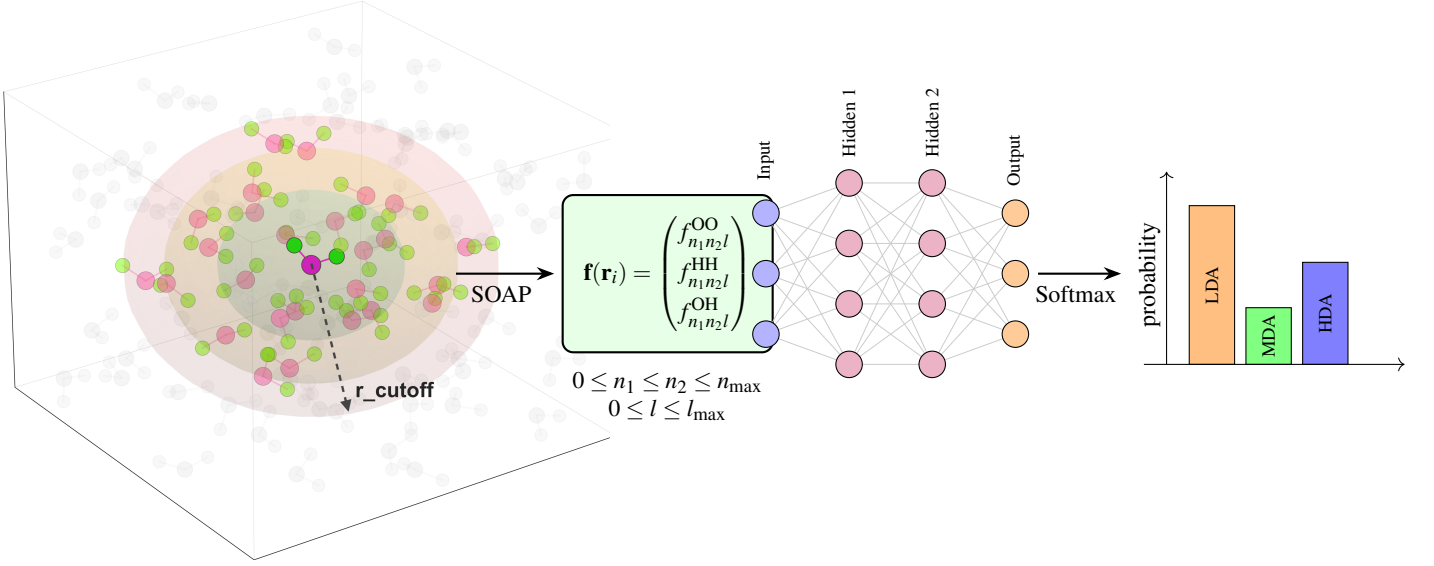

This section outlines the methodologies and tools employed to develop a model for classifying amorphous ices. The approach used in the work is schematically presented in Fig.~\ref{fig:combined_pipeline}. Subsection A provides details on the molecular dynamics simulations. Subsection B briefly introduces the principles underlying the SOAP descriptor. Subsection C describes the neural network architectures and the evaluation metrics used.

\subsection{\label{sec:MD_details} Molecular dynamics details}
The data set was obtained using molecular dynamics approaches. 
The initial amorphous phases LDA and HDA were created according to known recipes, described, for example, in previous work~\cite{garkul2022molecular, wong2015pressure}. 
LDA was obtained by ultra-fast isobaric cooling of liquid water to 77~K at 1~bar in an NPT assembly at a rate of 10~K/ns. To obtain HDA, hexagonal ice served as the initial structure, which was isothermally compressed at temperature of 77~K to a pressure of 2~GPa in an NPT-ensemble, and then unloaded to a lower pressure. The structure for MDA was taken from work of Rosu-Finsen~et~al.~\cite{rosu2023medium}, which also provides a detailed algorithm for obtaining MDA from Ih in simulations by randomly displacing atomic layers. The number of molecules in all amorphous systems is 2880.

Then several MD calculations with the TIP4P/Ice potential~\cite{abascal2005potential} were performed using the LAMMPS software package~\cite{lammps2022paper}. Those computations include compression, decompression, heating, cooling, and relaxation processes performed in $NVT$ or $NPT$ ensembles under different conditions and parameters. The integration time step is 2~fs. In each simulation, the O-H bonds are additionally constrained using the SHAKE algorithm~\cite{ryckaert1977numerical}. Long-range electrostatic interactions are treated using the Particle-Particle-Particle-Mesh algorithm~\cite{hockney2021computer} with an accuracy of $10^{-5}$.
The cutoff radii for the Coulomb and Lennard-Jones interactions are $r_{c, Coul}$~=~10~\AA~and $r_{c, LJ}$~=~12~\AA, respectively.

\subsection{\label{sec:descr}Local atomic structure description}

The SOAP descriptor is constructed as follows: for each central oxygen atom, all neighboring atoms within a spherical region of radius $r_{\text{cut}}$ are considered. The local atomic density is represented as a sum of Gaussian kernels centered on the positions of neighbors:
\begin{equation}
\rho(\mathbf{r}) = \sum_{j} \exp\left(-\frac{|\mathbf{r} - \mathbf{r}_j|^2}{2\sigma^2}\right),
\end{equation}
where the summation is over all neighboring atoms. The parameter $\sigma$ determines the width of the Gaussian kernel and acts as a regularization parameter, controlling the degree of smoothing of the atomic density. The choice of $\sigma$ establishes a compromise between sensitivity to the details of the atomic environment and robustness to thermal fluctuations and positional uncertainty of the atoms.

The resulting continuous density function is then expanded in a basis set of radial functions and spherical harmonics:
\begin{equation}
\rho(\mathbf{r}) \approx \sum_{n=0}^{n_{\max}} \sum_{l=0}^{l_{\max}} \sum_{m=-l}^{l} c_{nlm} \, g_n(r) \, Y_{lm}(\theta, \phi),
\end{equation}
where $g_n(r)$ are radial basis functions (typically orthogonal polynomials), $Y_{lm}(\theta,\phi)$ are spherical harmonics, and $c_{nlm}$ are expansion coefficients. The SOAP descriptor is built from the power spectrum of these coefficients, which is rotationally invariant:
\begin{equation}
f_{n_1 n_2 l} = \sum_{m=-l}^{l} c_{n_1 l m}^* \, c_{n_2 l m}.
\end{equation}
Here, the indices $n_1$ and $n_2$ correspond to the radial channels, while $l$ corresponds to the angular momentum channel. A brief physical interpretation of various $n$ and $l$ values is given in Table~\ref{tab:soap_indices}. Thus, SOAP provides a compact yet informative description of the local neighborhood that is sensitive to both the radial distribution of neighbors (density, coordination numbers) and their angular organization (tetrahedrality, ring statistics, degree of disorder). 

\begin{table*}
\centering
\caption{Physical interpretation of SOAP descriptor indices $n$ and $l$.}
\label{tab:soap_indices}
\begin{tabular}{|c|c|c|}
\hline
\textbf{Index} & \textbf{Physical meaning} & \textbf{Description} \\
\hline
\multicolumn{3}{|c|}{\textbf{Radial index $n$}} \\
\hline
0 & Coordination number & Total number of neighbors \\
1 & Short-range distribution & First neighbors, $\sim2.5$--$3.5$~\AA \\
2, 3 & Medium-range order & Second and third coordination spheres, $\sim4$--$6$~\AA \\
$\geq 4$ & Fine radial structure & Higher-resolution radial features \\
\hline
\multicolumn{3}{|c|}{\textbf{Angular index $l$}} \\
\hline
0 & Isotropic component & Coordination number (spherically symmetric) \\
1 & Dipole asymmetry & Bond directionality \\
2 & Quadrupole component & Deviations from spherical symmetry \\
3 & Tetrahedral symmetry & Key for ices; ideal tetrahedron gives strong $l=3$ signal \\
4, 5, 6 & Complex angular moduli & Sensitive to pentagonal/hexagonal rings and H-bond network distortions \\
7, 8 & High-angle features & Local disorder and deviations from ideal symmetry \\
\hline
\end{tabular}
\end{table*}

The dimension of the descriptor is determined by the parameters $n_{\max}$ and $l_{\max}$. In this work, two different versions of the SOAP descriptor were used. The first vector was constructed using only oxygen atoms, with the $n_{max}$ and $l_{max}$ parameters equal to 8. In this case, the descriptor dimension is 324 components. In the second case, we considered both oxygen and hydrogen atoms, but the $n_{max}$ and $l_{max}$ parameters were reduced to 6 to conserve computational resources. This vector contains 546 components. In both cases, the cutoff radius $r_{cut}~=~6.0$~\AA\ was chosen as a compromise value, including the first and second coordination spheres of oxygen in amorphous ice but excluding longer-range correlations that are insignificant or noisy. And the parameter $\sigma~=~0.5$~\AA\ is typical for systems with hydrogen bonds~\cite{bartok2013representing, monserrat2020liquid}.

The descriptor calculation was implemented using the DScribe library~\cite{laakso2023updates, himanen2020dscribe} for Python3.


\subsection{\label{sec:ML}Machine Learning approach}
\subsubsection{\label{sec:models}Model}

We use a classic neural network model for classification problems. It receives a descriptor vector as an input, then it transforms several times via linear transformation and activation function to finally form an output vector (see Fig.~\ref{fig:combined_pipeline}). The model contains two hidden layers with 600 neurons each. The activation function for the hidden layers is the hyperbolic tangent, and for the output layer the identity function. The structure of model, its training algorithm, and other parameters were chosen once and fixed during a comparative analysis in terms of final classification precision. 

To train and test the model results, the generated sample is divided into three parts: a training sample, a validation sample, and a test sample. The division occurs with random mixing in a 60:20:20 ratio, with each of the samples being balanced, meaning that there is an equal number of elements from each of the water phases being considered. The training sample is used directly to optimize the model parameters. The validation sample is also used in the training. After each training step, called an epoch, the error on the validation sample is calculated. This allows avoiding overtraining, which is detected by the error growth on this sample. In addition, a dropout method with a probability of 0.2 and early stopping if the loss function on the validation data set did not improve within 10 epochs were used to prevent overfitting. The test sample is used to calculate metrics that reflect the quality of the trained model. We used an initial learning rate of 0.001 with the Adam optimizer~\cite{kingma2014adam}. 
The model was constructed via PyTorch~\cite{ansel2024pytorch} framework for Python.

The study utilizes two types of NN models trained on two (LDA and HDA) and three (LDA, HDA, and MDA) types of amorphous structures, referred to as $M_2$ and $M_3$, respectively. 
For each structure, the training set contains 12~000 MD configurations with different T and P parameters. For LDA, the temperature range is from 20 to 120~K, and the pressure range is from 0 to 0.15~GPa. For MDA, the temperature range is from 40 to 120 K, and the pressure range is from 0.1 to 0.4~GPa. For HDA, the temperature range is from 40 to 120~K, the pressure range is from 0.3 to 0.6~GPa

\subsubsection{\label{sec:level3}Metrics}
To assess the performance of the neural network classifier, we employ the accuracy metric, defined as the ratio of correctly classified samples to the total number of samples in the test set. Since our data set is balanced, accuracy provides a reliable and interpretable measure of overall classification performance. The model is trained by minimizing the cross-entropy loss function, which is the standard choice for multi-class classification problems.

For quantitative assessment of structural identification at the atomic level, the trained model provides not only discrete class labels but also prediction probabilities. The final layer of the network produces logits $z_i$ -- real-valued scores representing the raw confidence for each class $i$. These logits are converted into probabilities using the softmax function:

\begin{equation}
p_i = \frac{e^{z_i}}{\sum_{j=1}^{K} e^{z_j}},
\label{eq:softmax}
\end{equation}

where $K$ is the number of classes. The resulting probability vector $\mathbf{p} = (p_1, p_2, \dots, p_K)$ satisfies $\sum_{i=1}^{K} p_i = 1$, with $p_i$ representing the probability that a given atom belongs to class $i$. The class with the highest probability is assigned as the final label, and the corresponding probability value serves as a quantitative measure of prediction confidence.

Feature importance is quantified using permutation importance~\cite{altmann2010permutation}, a model-agnostic method that measures the decrease in classification accuracy when a feature's values are randomly shuffled. For each feature $j$, we compute:
\begin{equation}
I_j = A_{\text{base}} - \frac{1}{R}\sum_{r=1}^{R} A_{\text{perm}}^{(j,r)},
\end{equation}
where $A_{\text{base}}$ is the baseline accuracy on the test set, $A_{\text{perm}}^{(j,r)}$ is the accuracy after permuting feature $j$ in repetition $r$ ($R = 10$), and $I_j$ represents the importance of feature $j$. Larger $I_j$ indicates greater discriminative power. Standard deviations are estimated from the $R$ permutations to assess robustness.

\section{\label{sec:results}Results}
\subsection{\label{sec:model_learning} Training NN-models and feature importance analysis}

\begin{figure}[ht]
\centering
    \begin{minipage}{0.45\linewidth}
    \centering
    \includegraphics[trim=1 1 90 1, clip, width=0.99\textwidth]{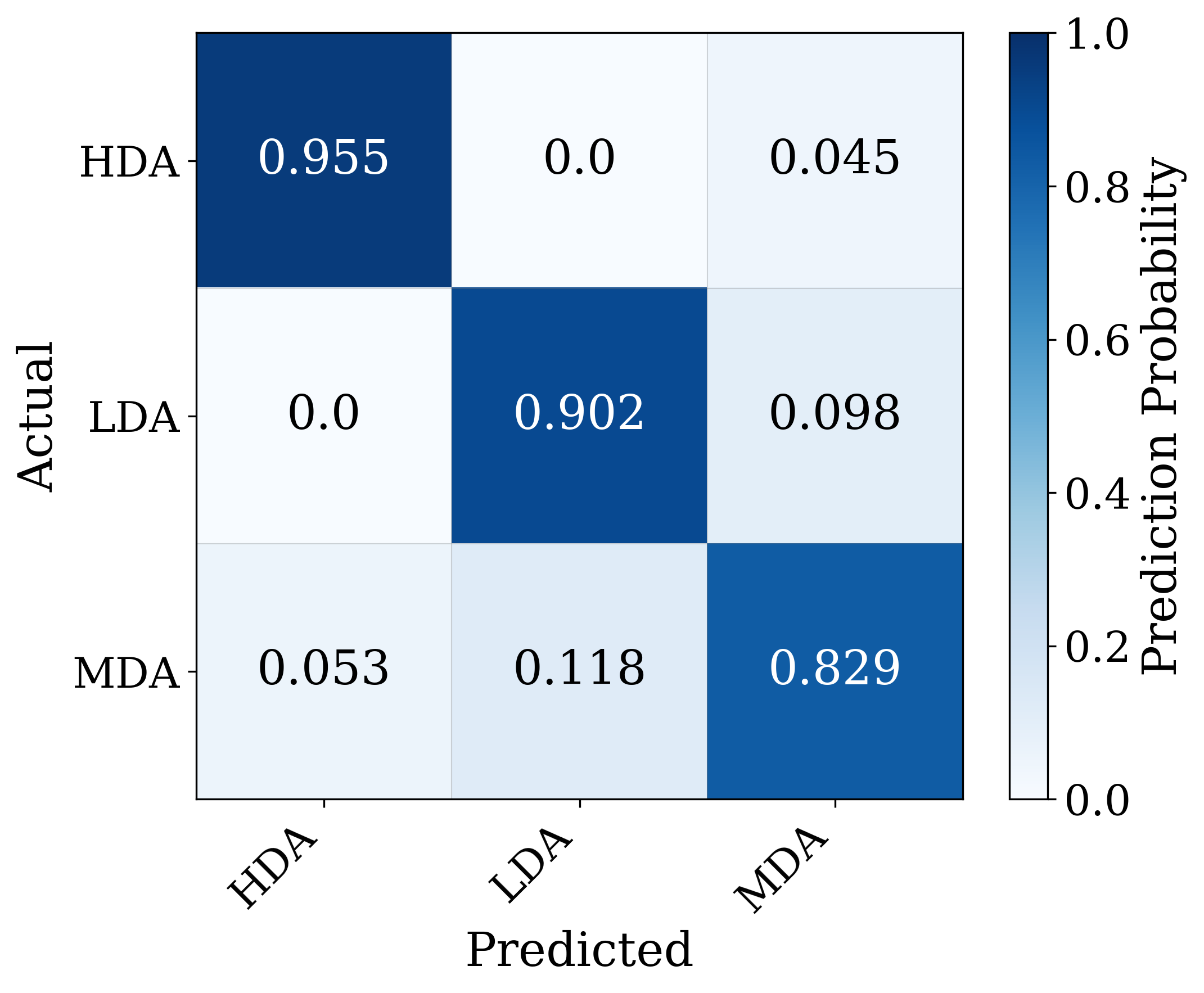} a)
    \end{minipage}
    \hfill
    \begin{minipage}{0.53\linewidth}
    \centering
    \includegraphics[trim=1 1 1 1, clip, width=0.99\linewidth]{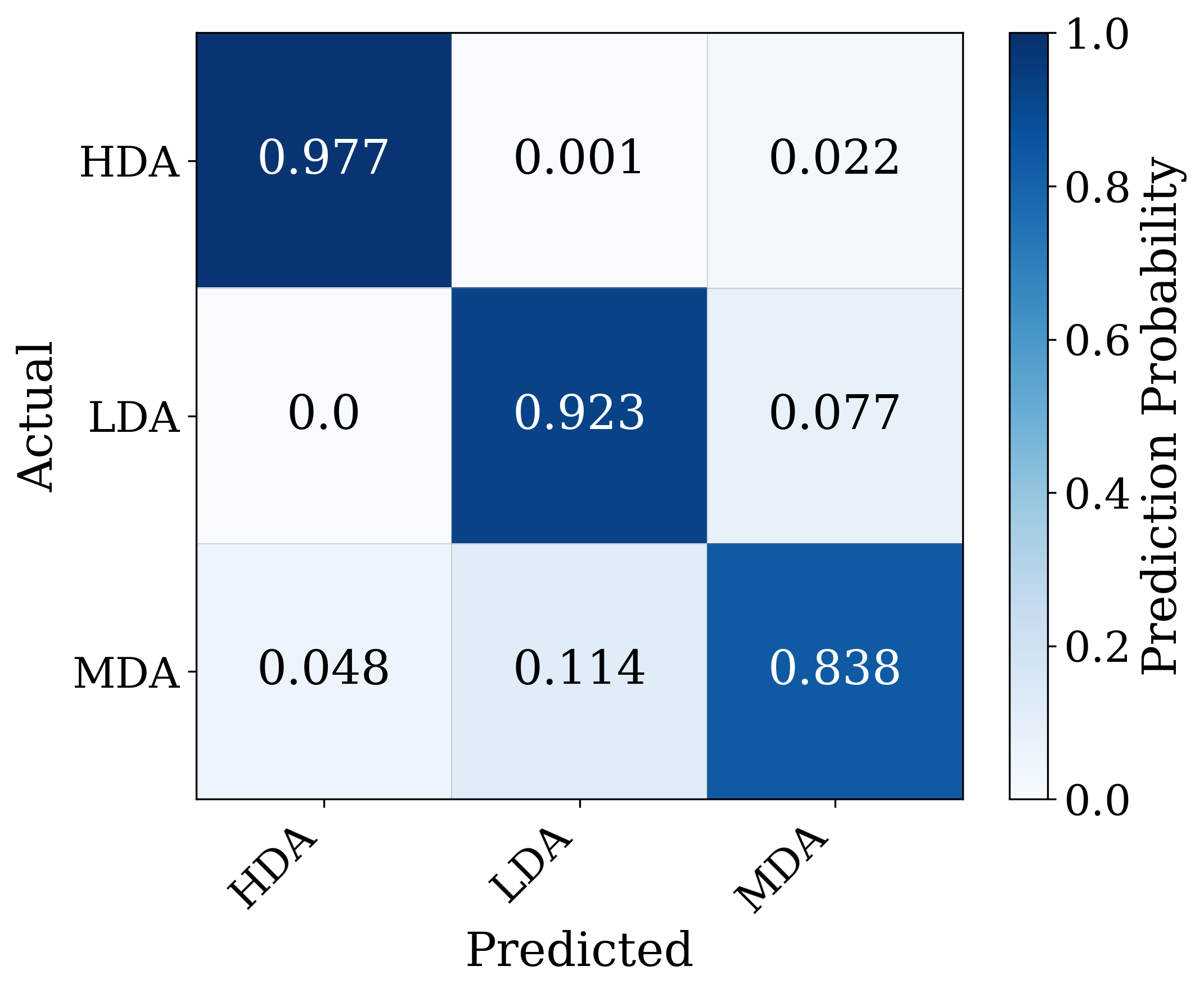} b)
    \end{minipage}
\caption{Confusion matrices for a)~$M_3(O)$ and b)~$M_3(O, H)$ models. The vertical axis is the actual phase of the structure, the horizontal axis is the predicted phase of the structure. Each cell of the matrix indicates the probability with which the model classifies a particular structure as a system with the corresponding phase.}
\label{fig:CM_soap}
\end{figure}

We first trained a SOAP-based model on three amorphous structures and analyzed feature importance using permutation importance and performed principal component analysis (PCA) on the descriptor space to identify structural fingerprints distinguishing HDA, LDA and MDA. $M_3(O)$ considers only oxygen atoms when calculating descriptors, while $M_3(O, H)$ considers both oxygen and hydrogen. The confusion matrix is shown in Figure~\ref{fig:CM_soap}. The classification accuracies of these models differ slightly, amounting to 0.8953 and 0.9005. However, adding information about hydrogen atoms leads to a redistribution in the ranking of important features. Table~\ref{tab:pairwise_top7} presents the most important features for pairwise discrimination of amorphous ices (see also Supplementary Fig.~S1 online).

\begin{table}[h!]
\centering
\caption{Top-7 most important features for pairwise phase discrimination in $M_3(O)$ and $M_3(O, H)$ models. The values of the permutation importance of the feature are given in brackets.}
\label{tab:pairwise_top7}
\begin{tabular}{|c|c|c|}
\hline
\multicolumn{3}{|c|}{\textbf{$M_3(O)$}} \\
\hline
\textbf{HDA vs LDA} & \textbf{LDA vs MDA} & \textbf{HDA vs MDA}\\
\hline
OO\_008 \textcolor{gray}{(0.063)} & OO\_000 \textcolor{gray}{(0.074)} & OO\_000 \textcolor{gray}{(0.082)} \\
OO\_000 \textcolor{gray}{(0.059)} & OO\_357 \textcolor{gray}{(0.045)} & OO\_267 \textcolor{gray}{(0.028)} \\
OO\_022 \textcolor{gray}{(0.048)} & OO\_008 \textcolor{gray}{(0.036)} & OO\_457 \textcolor{gray}{(0.027)} \\
OO\_357 \textcolor{gray}{(0.048)} & OO\_265 \textcolor{gray}{(0.033)} & OO\_148 \textcolor{gray}{(0.025)} \\
OO\_007 \textcolor{gray}{(0.039)} & OO\_267 \textcolor{gray}{(0.031)} & OO\_228 \textcolor{gray}{(0.020)} \\
OO\_265 \textcolor{gray}{(0.039)} & OO\_457 \textcolor{gray}{(0.029)} & OO\_252 \textcolor{gray}{(0.020)} \\
OO\_015 \textcolor{gray}{(0.037)} & OO\_165 \textcolor{gray}{(0.022)} & OO\_245 \textcolor{gray}{(0.019)} \\
\hline
\multicolumn{3}{|c|}{\textbf{$M_3(O, H)$}} \\
\hline
\textbf{HDA vs LDA} & \textbf{LDA vs MDA} & \textbf{HDA vs MDA}\\
\hline
OH\_256 \textcolor{gray}{(0.039)} & HH\_000 \textcolor{gray}{(0.018)} & OO\_226 \textcolor{gray}{(0.030)} \\
OH\_254 \textcolor{gray}{(0.030)} & OO\_355 \textcolor{gray}{(0.017)} & OH\_226 \textcolor{gray}{(0.026)} \\
OH\_255 \textcolor{gray}{(0.029)} & OO\_225 \textcolor{gray}{(0.011)} & OO\_136 \textcolor{gray}{(0.022)} \\
HH\_050 \textcolor{gray}{(0.022)} & OH\_135 \textcolor{gray}{(0.011)} & OH\_135 \textcolor{gray}{(0.018)} \\
OH\_556 \textcolor{gray}{(0.018)} & OO\_003 \textcolor{gray}{(0.011)} & OO\_255 \textcolor{gray}{(0.017)} \\
OO\_555 \textcolor{gray}{(0.018)} & OH\_044 \textcolor{gray}{(0.010)} & OH\_220 \textcolor{gray}{(0.015)} \\
HH\_046 \textcolor{gray}{(0.017)} & OO\_136 \textcolor{gray}{(0.010)} & OO\_056 \textcolor{gray}{(0.014)} \\
\hline
\end{tabular}
\\[5pt]
\footnotesize{Note: For $M_3(O)$, ``OO\_$xyz$'' denotes $(n_1=x, n_2=y, l=z)$; for $M_3(O, H)$, ``OO\_$xyz$'', ``HH\_$xyz$'', and ``OH\_$xyz$'' specify the atomic pair types.}
\end{table}

In the $M_3(O)$ model, oxygen coordination number (OO\_000) and high-$l$ angular modes (OO\_$n_1n_28$, OO\_$n_1n_27$) dominate all comparisons, indicating that density and tetrahedral distortion are the primary discriminators. Meanwhile, the $M_3(O, H)$ model reveals a more nuanced picture: HDA–LDA discrimination is dominated by OH features (OH\_256, OH\_254, OH\_255) with high angular moments ($l~=~4,~5,~6$), demonstrating that hydrogen-bond orientation is the key distinguishing factor. The LDA–MDA distinction is mainly based on the hydrogen packing density in the first coordination sphere (HH\_000). Next in importance are oxygen–oxygen correlations (OO\_355, OO\_225), which describe distortions of the oxygen sublattice. For HDA versus MDA, the main features are OO\_226, OH\_226 and OO\_136, which probe the medium-range order of both the oxygen network and hydrogen bonds (second coordination shell). Thus, despite similar densities, HDA and MDA differ in the spatial arrangement of the oxygen sublattice, with MDA retaining more tetrahedral character inherited from LDA. 

\begin{figure*}[ht]
    \begin{minipage}[h]{0.99\linewidth}
    \centering
    \includegraphics[trim=1 1 1 1, clip, width=\textwidth]{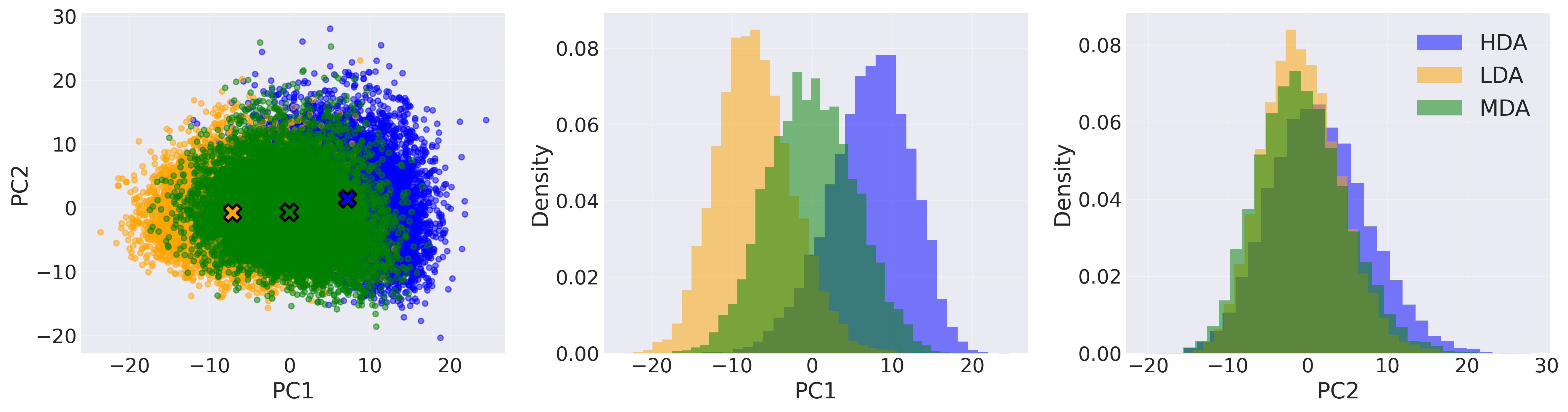}
    \end{minipage}
    \hfill
    \begin{minipage}[h]{0.99\linewidth}
    \centering
    \includegraphics[trim=1 1 1 1, clip, width=\linewidth]{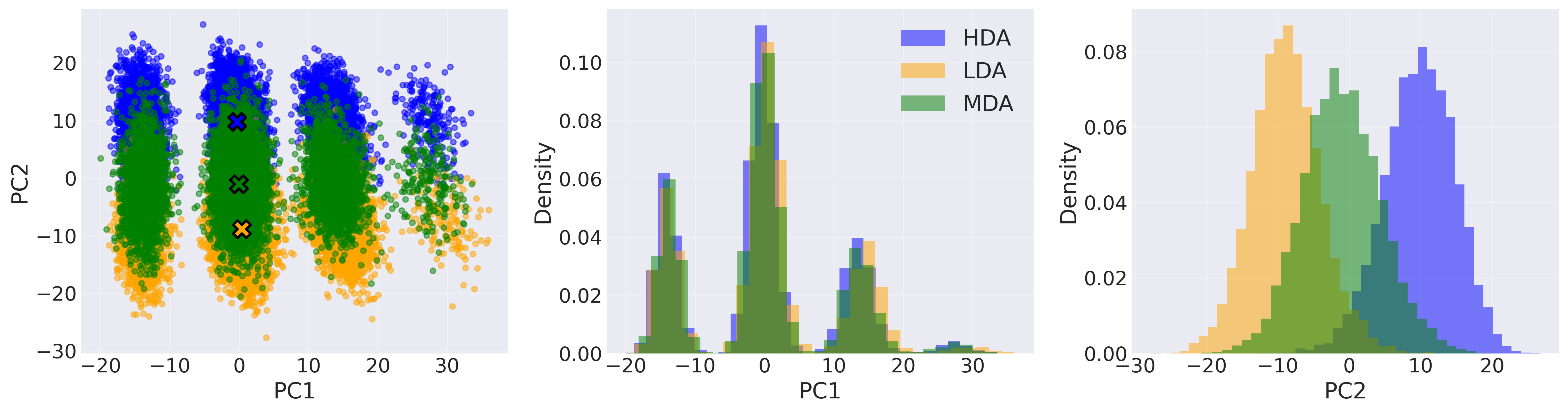}
    \end{minipage}
\caption{Distributions of sample data in principal component space. The top panel shows the results for model $M_3(O)$, the bottom panel shows the results for model $M_3(O, H)$.}
\label{fig:pca}
\end{figure*}

Principal Component Analysis (PCA) was performed on the feature space of both models. Table~II in Supplementary Information shows the top-10 features in each model by contribution to the main components. Figure~\ref{fig:pca} illustrates the phase separation in the first two principal components, and Table~\ref{tab:pca_results} summarizes the key metrics.

\begin{table}[h]
\centering
\caption{PCA results for $M_3(O)$ and $M_3(O, H)$ models}
\label{tab:pca_results}
\begin{tabular}{|c|c|c|}
\hline
\textbf{Metric} & \textbf{$M_3(O)$} & \textbf{$M_3(O, H)$} \\
\hline
PC1 variance & 18.6\% & 19.7\% \\
PC2 variance & 9.8\% & 15.5\% \\
PC1+PC2 variance & 28.4\% & 35.2\% \\
Components for 95\% variance & 34 & 66 \\
\hline
HDA centroid & (7.19, 1.45) & (-0.30, 9.82) \\
LDA centroid & (-7.16, -0.78) & (0.37, -8.83) \\
MDA centroid & (-0.07, -0.67) & (-0.07, -1.05) \\
\hline
HDA–LDA distance & 14.53 & 18.67 \\
HDA–MDA distance & 7.56 & 10.88 \\
LDA–MDA distance & 7.09 & 7.79 \\
\hline
PC1 separation accuracy & 0.6762 & 0.3870 \\
PC2 separation accuracy & 0.3573 & 0.7566 \\
\hline
\end{tabular}
\end{table}

The first two principal components in the $M_3(O)$ model explain $28.4\%$ of the total variance ($18.6\%$ for PC1 and $9.8\%$ for PC2). PC1 captures the dominant mode of variation, which is primarily associated with local density, as evidenced by the separation accuracy of $0.676$ when used as a single-feature linear classifier. In contrast, PC2 achieves only $0.387$ accuracy, indicating that orientational information is less pronounced when hydrogen atoms are omitted. The centroids of HDA and LDA are well-separated along PC1 ($7.19$ vs $-7.16$), while MDA occupies an intermediate position ($-0.07$), consistent with its intermediate density.

The inclusion of hydrogen atoms increases the total variance explained by the first two components to $35.2\%$ ($19.7\%$ for PC1 and $15.5\%$ for PC2). The first principal component primarily captures intra-phase structural diversity, as evidenced by the broad distribution of points along this axis for each individual phase. The second principal component, while explaining a smaller fraction of variance, provides clear separation between phases: HDA environments occupy high PC2 values, LDA environments low PC2 values, and MDA environments lie in between. This interpretation is quantitatively confirmed by the separation accuracies: PC1 achieves only $0.357$, whereas PC2 reaches $0.757$. This striking reversal compared to the $M_3(O)$ model demonstrates that the inclusion of hydrogen atoms shifts the most discriminative information from the first to the second principal component. PC2 encodes the key structural features that discriminate between amorphous ice phases — namely, hydrogen-bond orientation and medium-range order ($4$-$6$~\AA) -- while PC1 reflects the inherent structural heterogeneity within each phase.

\begin{figure*}[!ht]
    \begin{minipage}[h]{0.49\linewidth}
    \centering
    \includegraphics[trim=1 1 500 7, clip, width=\textwidth]{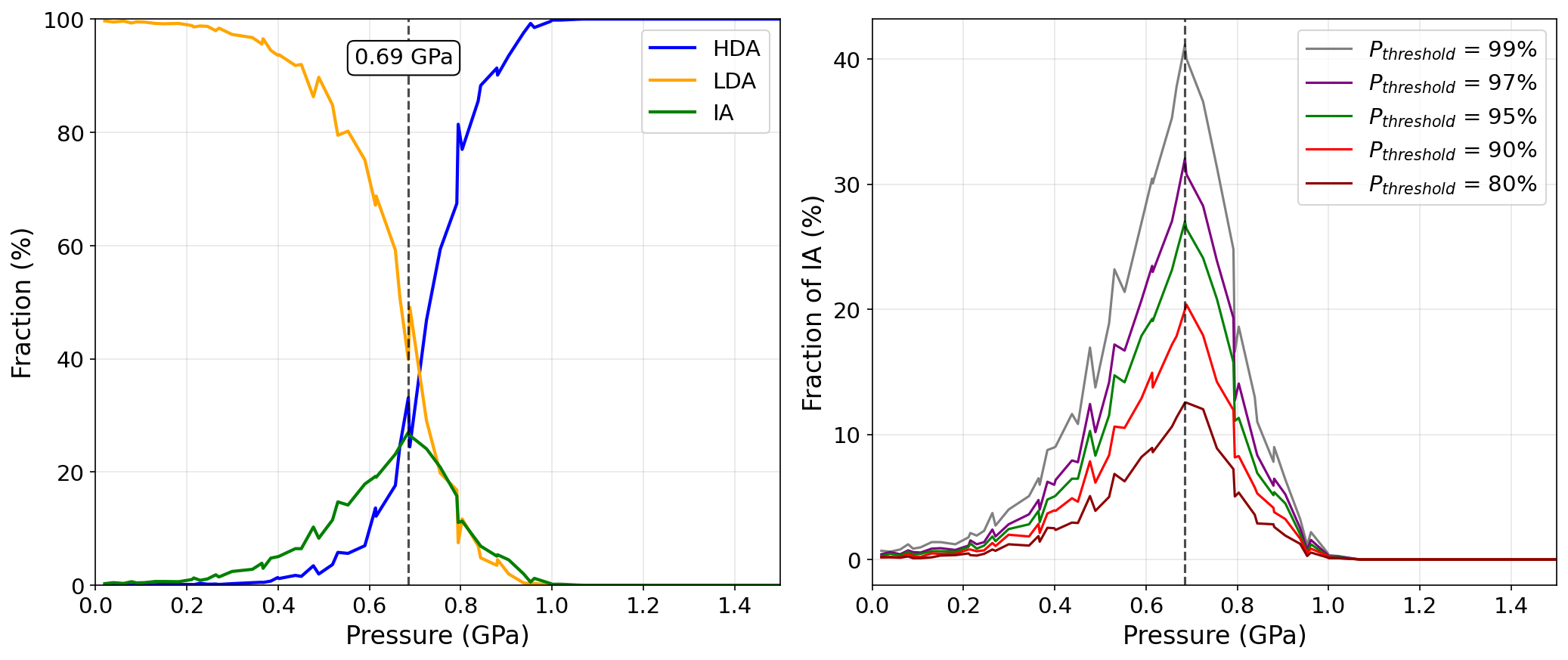} a)
    \end{minipage}
    \hfill
    \begin{minipage}[h]{0.49\linewidth}
    \centering
    \includegraphics[trim=1 0 5 0, clip, width=0.98\linewidth]{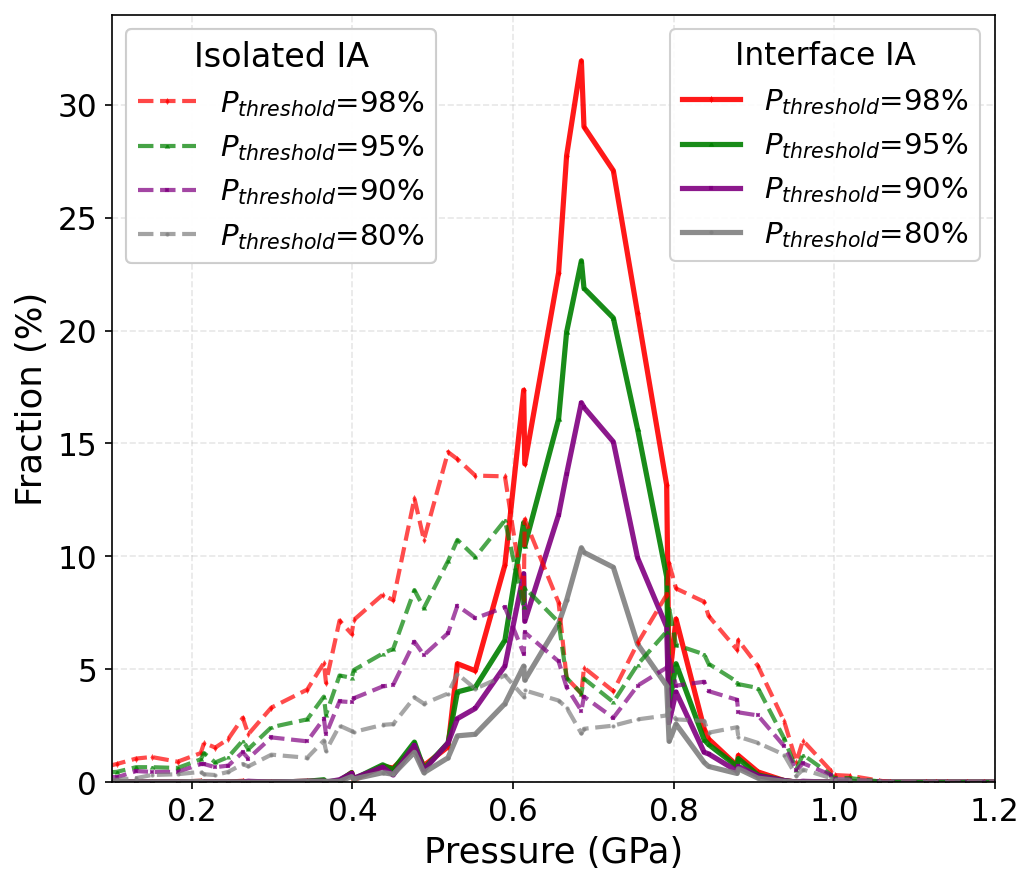} b)
    \end{minipage}
\caption{ a)~The dependence of the fraction of LDA, HDA, and IA phases on pressure during isothermal compression of LDA ($P_{threshold}~=~0.95$). b)~The fraction of the IA phase for different confidence thresholds from 0.80 to 0.98.}
\label{fig:phase_evolution}
\end{figure*}

\begin{figure*}[ht]
	\centering
	\includegraphics[trim=1 1 1 1, clip, width=0.99\textwidth]{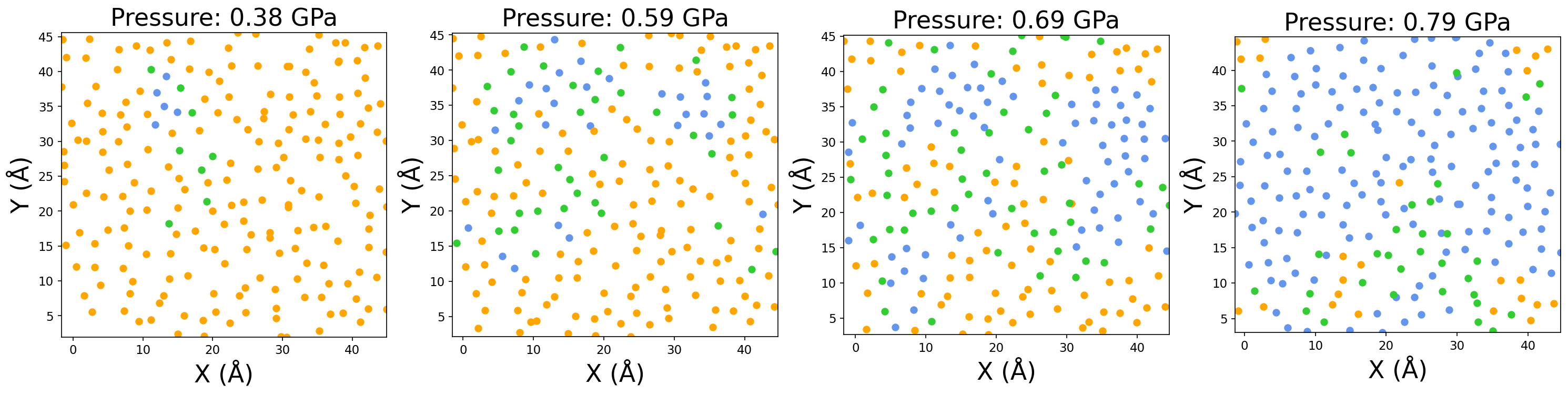}
\caption{Snapshots of 3.0 \AA-thick slices of the computational cell along the plane $z$~=~22~\AA~at moments corresponding to pressures of 0.38, 0.59, 0.69 and 0.79~GPa. Oxygen atoms in molecules recognized by the model as HDA are colored blue, as LDA -- orange, and as IA -- green.}
\label{fig:slice}
\end{figure*}

Adding hydrogen atoms significantly improves phase separability: HDA–LDA distance increases by $28\%$, HDA-MDA by $44\%$, and LDA–MDA by $10\%$. However, more components are needed to explain $95\%$ of variance ($66$ vs $34$), reflecting the higher dimensionality and complexity introduced by hydrogen-related features.

To further validate the essential role of hydrogen-related features, we trained a separate $M_3(H\text{-}bond)$ model using only OH and HH descriptors, excluding all OO\_$n_1n_2l$ contributions. Remarkably, this reduced model achieves an accuracy of 90.92\%, which is even slightly higher than the accuracy of $M_3(O, H)$ model (see Supplementary Table~I). This indicates that all necessary information for phase discrimination is contained in hydrogen-related structural features (see Supplementary Table~III). PCA of the $M_3(H\text{-}bond)$ model shows that PC2 (which captures medium-range order) remains the primary discriminative axis, while the number of components required to explain 95\% of the variance decreases from 66 to 54, demonstrating that OO features contribute primarily to noise and dimensionality without adding discriminative power. These results provide compelling evidence that the spatial distribution and orientational correlations of hydrogen atoms are key structural features distinguishing amorphous ice phases.
Models based solely on oxygen atom positions account for density differences but overlook critical orientational and topological features.

It is worth noting that the separation between HDA and LDA in the PCA space becomes even more pronounced when MDA is excluded from the analysis. In the binary classifier $M_2(O,H)$ trained exclusively on HDA and LDA, the distance between the phase centroids increases by approximately 25\% compared to the three-class model $M_3(O,H)$, reflecting the fact that PCA in the binary case is optimized solely to capture the variance between HDA and LDA, without needing to also represent the intermediate MDA structures.

To avoid potential ambiguity introduced by the MDA phase, which occupies an intermediate position between HDA and LDA in PCA space (Fig.~\ref{fig:pca}), subsequent analysis of the structural evolution during phase transitions was performed using binary classifiers trained exclusively on HDA and LDA samples. These models achieve near-perfect accuracy: 99.78\% for $M_2(O)$ and 99.83\% for $M_2(O,H)$. Importantly, the feature importance ranking for HDA-LDA discrimination remains qualitatively unchanged compared to the three-class model, confirming that the key discriminant features identified in Table~\ref{tab:pairwise_top7} are robust and independent of the presence of intermediate phases. To identify possible intermediate amorphous phases, we will consider the prediction probability. If the probability with which the model will classify the structure as an LDA or HDA phase is below the threshold value, then we will consider the structure as an IA representative.

\subsection{\label{sec:PIT_analisys}Analysis of pressure-induced LDA~$\to$~HDA transformation}

To investigate the microscopic mechanism of the LDA~$\to$~HDA transition, a series of molecular dynamics simulations were performed in an isothermal NPT ensemble. The initial LDA configuration was compressed at a rate of 0.1~GPa/ns at a fixed temperature $T$~=~77~K. During the compression process, the atomic configurations were preserved every 100~000 integration steps (which corresponds to 20~ps), forming a trajectory of $100$ successive frames covering the full pressure range from atmospheric to 2.0~GPa. For each frame, SOAP local environment descriptors were calculated, which were subsequently used to analyze the structural evolution.

The local structure of each molecule was classified using the trained neural network models described in section~\ref{sec:models}. The models returns the probabilities of a molecule belonging to the LDA and HDA phases. To identify molecules with an intermediate amorphous (IA) structure, a confidence threshold $P_{threshold}$ is introduced. A molecule is classified as confidently belonging to LDA or HDA if the corresponding probability exceeds the threshold; otherwise, it is classified as an IA (uncertain) state. 

In the main text, we present the results obtained with the $M_2(O,H)$ model, which includes both oxygen and hydrogen atoms in the SOAP descriptor. This model provides a more complete description of the local environment, particularly capturing orientational information essential for distinguishing LDA and HDA (see Section~\ref{sec:model_learning}). Qualitatively similar results were obtained with the oxygen-only $M_2(O)$ model; these are presented in the Supplementary Material (Figs.~S2–S3) and show no significant differences in the overall transformation behavior.

\subsubsection{Redistribution of amorphous states}
Analysis of macroscopic characteristics, in particular the dependence of density on pressure, shows that the most intense structural rearrangement occurs in the pressure range of 0.5-0.8~GPa. According to our previous results, it is in this pressure range that the formation and subsequent intensive growth of HDA clusters in the LDA matrix occurs~\cite{garkul2022molecular}.

The figure~\ref{fig:phase_evolution}~(a) shows the dependence of the fractions of LDA, HDA, and IA phases on pressure at a confidence threshold of $P_{threshold}$~=~0.95.
At the initial stage of the transition (where the pressure is about 0.5 GPa), the overwhelming majority of molecules ($\sim 90\%$) are confidently identified as LDA, while the proportion of HDA is less than $5\%$. With increasing pressure, the proportion of HDA increases, but at the same time, a significant increase in the proportion of IA molecules is observed in the middle of the transition. Although oxygen atoms assigned to the IA phase can combine into small clusters of up to 20 molecules, significant cluster growth phenomena are not observed, as in the case of HDA (see next section~\ref{sec:cluster_analisys}). This suggests that IA arises at the LDA-HDA interface, not as a bulk phase.

Varying the confidence threshold $P_{threshold}$ allows one to control the rigor of the classification and evaluate the robustness of the obtained results. As the threshold increases, the proportion of molecules classified as IA increases, since the requirements for the classifier's confidence become more stringent (Fig.~\ref{fig:phase_evolution}~(b)). However, the qualitative behavior -- the evolution of the phase distribution along the compression trajectory -- is preserved regardless of the choice of $P_{threshold}$, which indicates the physical nature of the observed effects. The dependence of the maximum fraction of IA molecules on the threshold probability value (see Fig.~S7) is smooth, with no pronounced jumps that would indicate a ``natural'' threshold. That is, IA molecules do not form a clearly separable population. The choice of a threshold of 0.9-0.95 is reasonable, as lower thresholds lead to a loss of model sensitivity, while higher thresholds increase noise.

Notably, when the three-class classifiers $M_3(O)$ and $M_3(O,H)$ (trained on LDA, HDA, and MDA structures) are applied to the same compression trajectory, they classify a substantial fraction of molecules ($\sim$70\% at $P \approx 0.69$~GPa) as MDA, in agreement with the previous results~\cite{faure2024high}. This confirms that the IA molecules detected by the binary classifier indeed correspond to MDA-like structures.

For a detailed analysis of structural changes at the molecular level, we select key frames corresponding to pressures of 0.38, 0.5, 0.59, 0.69 and 0.8~GPa, which represent different stages of the transition: nucleation, growth, and coalescence of HDA clusters. For each frame, we visualize a slice of the simulation cell with a thickness of 3.0~\AA, showing oxygen atoms colored according to phase: LDA (orange), HDA (blue) and IA (green) (Fig.~\ref{fig:slice}). 
The spatial distribution clearly shows that IA molecules tend to form at the boundaries between LDA and HDA clusters. Isolated IA molecules also exist, but their fraction is significantly smaller. This observation allows us to refine the pressure dependence of the IA fraction by distinguishing between two populations: interfacial IA molecules, which have at least one LDA neighbor and one HDA neighbor within a radius of 3.0~\AA, and isolated IA molecules, which are surrounded exclusively by either LDA or HDA (Fig.~\ref{fig:phase_evolution}~(b)). The IA fraction increases primarily through the growth of the interfacial population. Varying the confidence threshold has only a minor effect on the fraction of isolated molecules.

\begin{figure}[ht]
	\centering
	\includegraphics[trim=1 3 0 3, clip, width=0.5\textwidth]{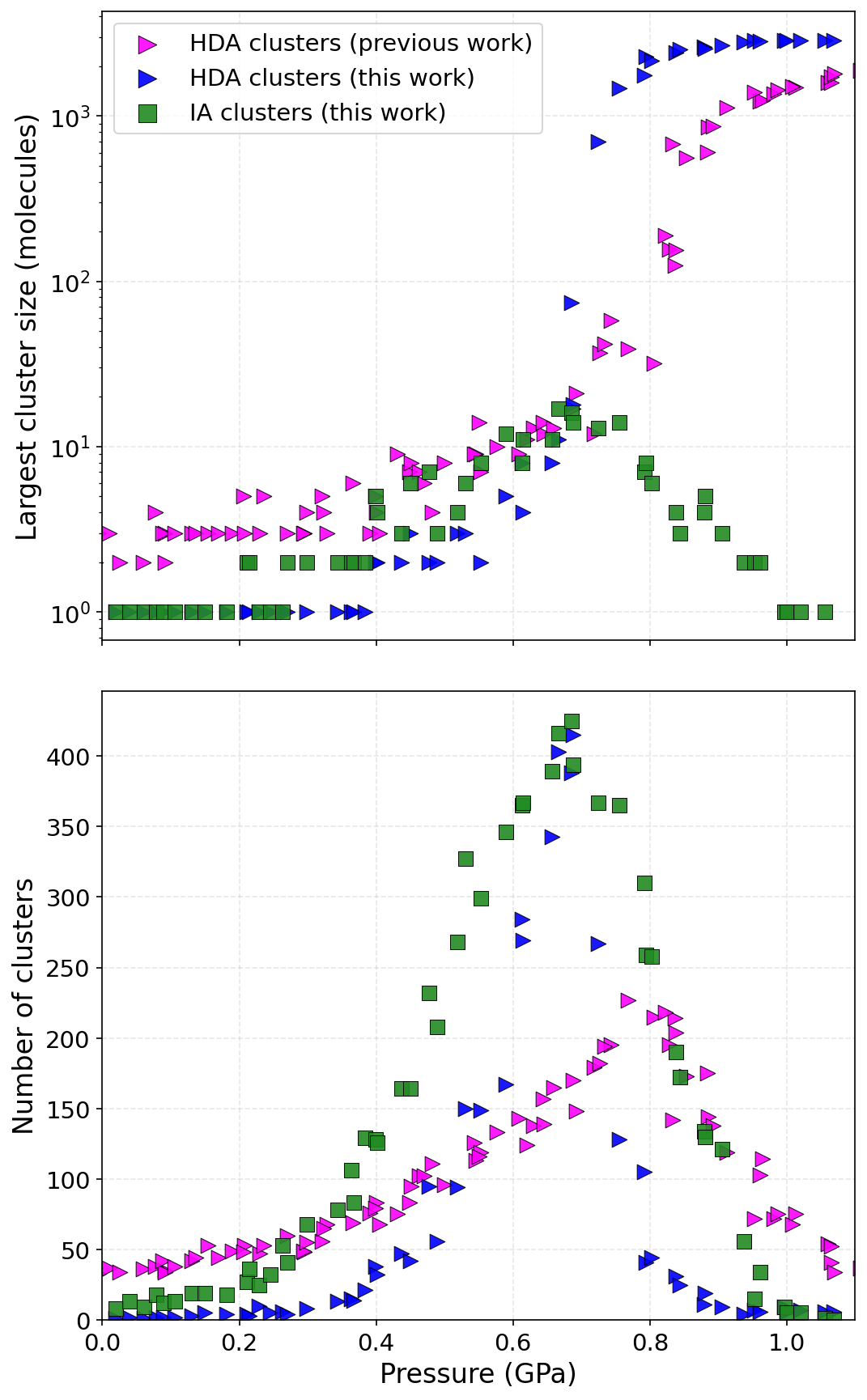}
    \caption{Evolution of HDA and IA clusters during pressure-induced polyamorphic transformation in this study for $M_2(O,H)$ at a confidence threshold of 0.95, and in the previous work of Garkul~\&~Stegailov ($N^{descr}\geq 11$)~\cite{garkul2022molecular}. a)~The pressure dependence of the size of the largest HDA and IA clusters in the system. b)~The pressure dependence of the number of HDA clusters.}
    \label{fig:hda_cluster_evolution}
\end{figure}

\subsubsection{\label{sec:cluster_analisys}Evolution of HDA and IA clusters}
To fully understand the microscopic mechanism of LDA to HDA conversion, we analyzed the nucleation and growth of HDA and IA clusters along the compression trajectory. Figure~\ref{fig:hda_cluster_evolution} shows the evolution of the largest cluster size and the number of clusters as a function of pressure. Clusters were defined as bound molecules within a radius of 3.0~\AA. The confidence threshold varied within a reasonable range (0.9-0.99), but no significant differences were observed (see Supplementary Fig.~S4 online). 

At low pressures ($P\leqslant0.5$~GPa), only small HDA clusters (less than 4 molecules) were present. The size of critical HDA nuclei before intense growth is approximately 5–10 molecules (this stage corresponds to a pressure of 0.6~GPa). This is consistent with our previous molecular dynamics study~\cite{garkul2022molecular} and experimental estimates of Tonauer~et~al.~\cite{tonauer2017high}. In this regime, the number of HDA clusters increases, indicating multiple independent nucleation events. Above $P~\thickapprox~0.6$~GPa, the size of the largest HDA cluster increases sharply, while the number of clusters reaches a maximum in the middle of the transition, where the pressure is 0.69~GPa, and then decreases due to coalescence. The transformation is complete above 1.0~GPa, where almost all HDA molecules belong to a single cluster.

In contrast, IA clusters exhibit a different behavior: clusters grow in size and quantity until the system reaches the transition pressure (0.69~GPa), and then a decline follows associated with a decrease in the total fraction of IA molecules.

\subsection{\label{sec:interface_analisys}Analysis of the LDA~||~HDA interface}
\subsubsection{\label{sec:interface_balanced}Equilibrium system with a flat interface}
In this section, we characterize the properties of an artificially created LDA~||~HDA interface. We model separate systems with low- and high-density amorphous ices at an equilibrium pressure of approximately 0.2~GPa and a temperature of 77~K. Constructing a flat interface between amorphous ices requires careful handling, especially in the initial stages. A detailed algorithm for creating an LDA~||~HDA~||~LDA system that includes two interfaces is described in the Supplementary Materials (Section~I). In total, the system contains 15~339 molecules. Relaxation of the system in the NVT ensemble after coupling the LDA and HDA takes 30~ns. This relaxation time is sufficient for the pressure and distribution of molecules at the interface according to the $M_2(O)$ model. However, the $M_2(O,H)$ model reflects longer relaxation processes (see Supplementary Fig.~S5 online). 
Since the temperature is significantly lower than the glass transition temperatures of both phases, we do not expect large-scale diffusion or boundary motion. Our modelling focuses on the equilibrium interface structure in the frozen glassy state, which is established through local relaxation. 

\begin{figure*}[ht]
    \begin{minipage}[h]{0.92\linewidth}
        \centering
        \includegraphics[trim=200 380 160 382, clip, width=0.99\textwidth]{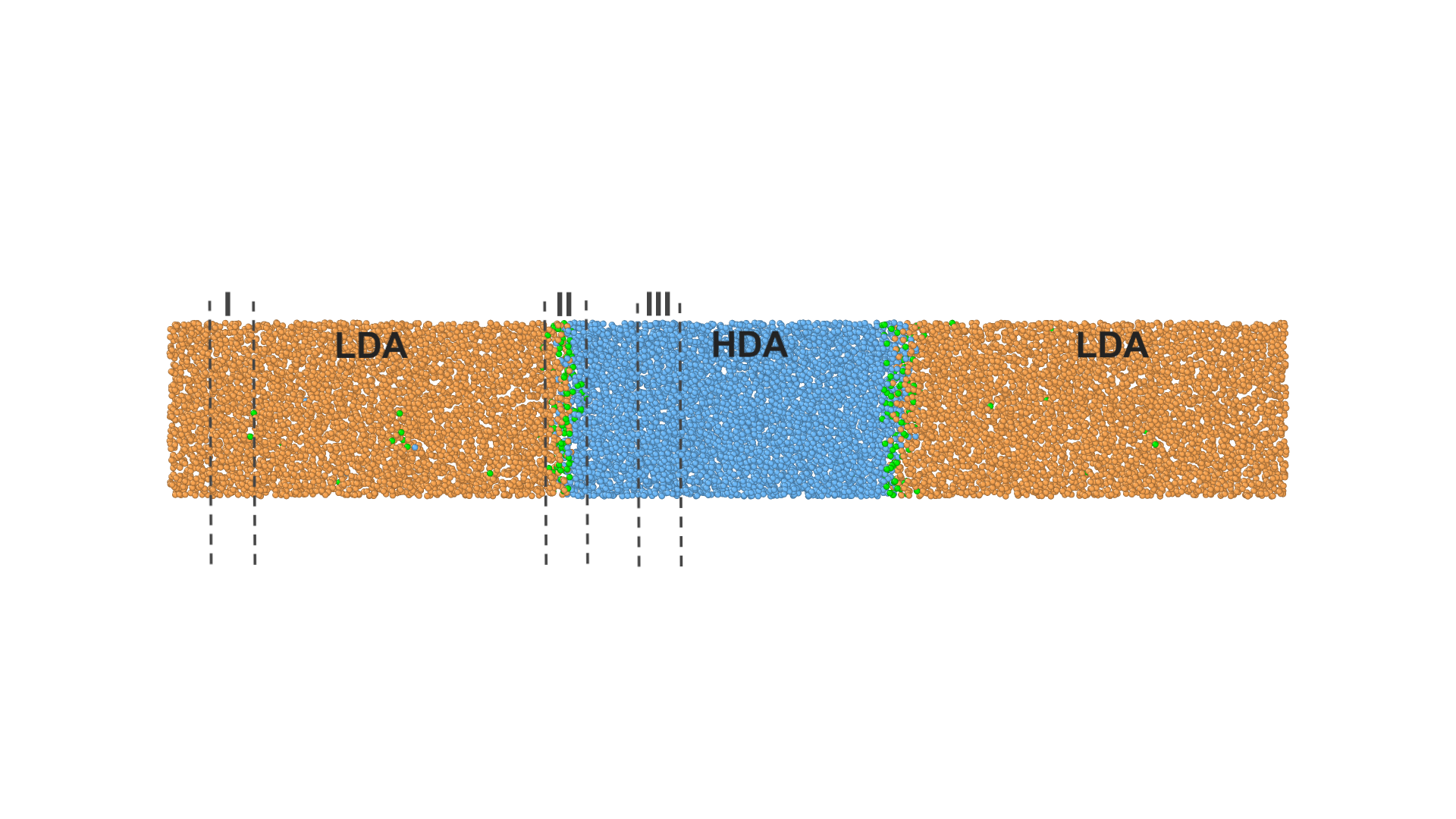}
    \end{minipage}
    \begin{minipage}[h]{\linewidth}
        \centering
        \includegraphics[trim=1 35 1 8, clip, width=0.99\linewidth]{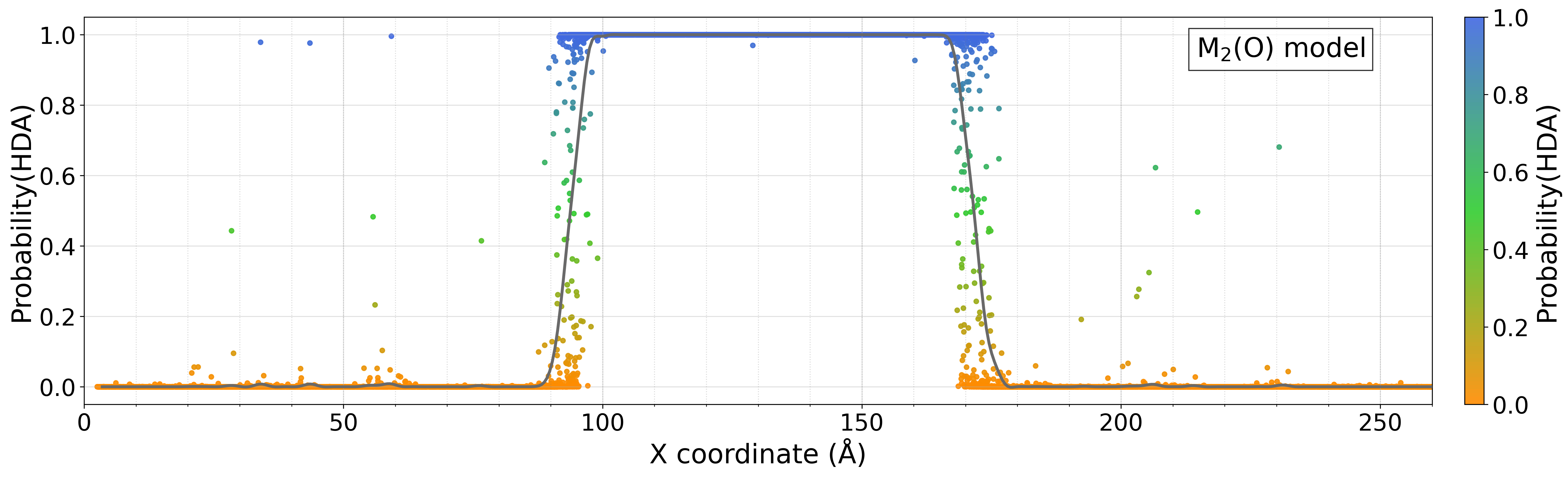}
    \end{minipage}
    \begin{minipage}[h]{\linewidth}
        \centering
        \includegraphics[trim=1 1 1 1, clip, width=0.99\textwidth]{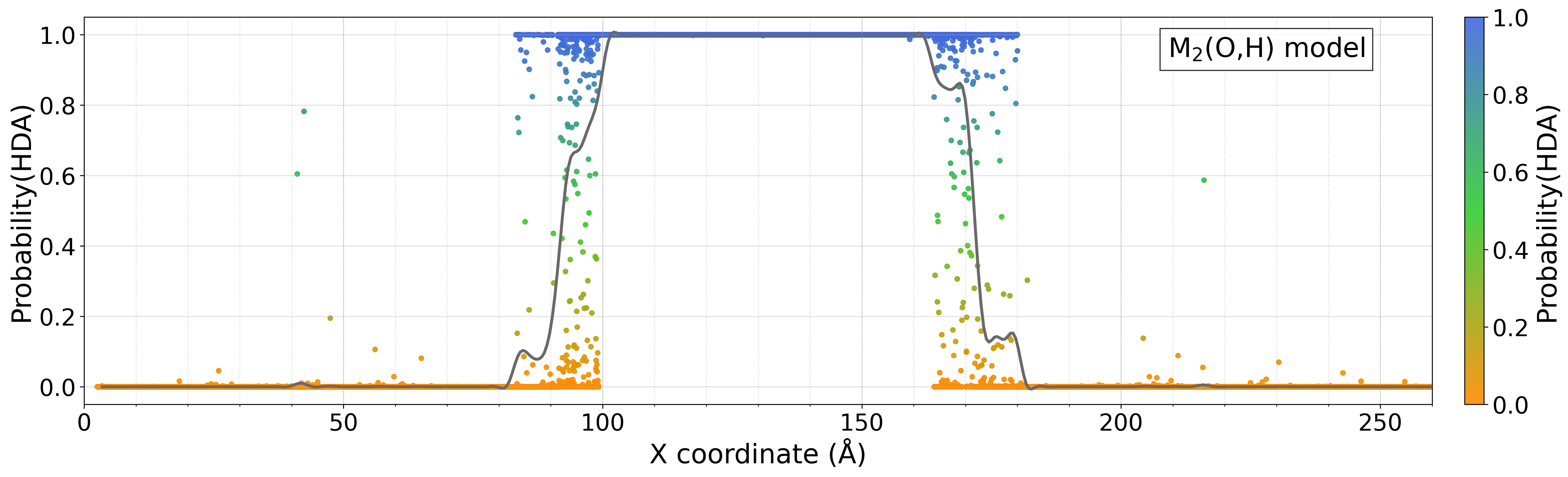}
    \end{minipage}
    \caption{Top panel: snapshot of the LDA~||~HDA~||~LDA system after relaxation at 77~K, pressure equal to 0.2~GPa. Oxygen atoms recognized by the $M_2(O)$ model as HDA are colored blue, LDA -- orange, and IA -- green. The confidence threshold is 0.95. Regions I, II and III, indicated by the dotted line, are selected for analysis in the PC space (see Fig.~\ref{fig:interface_pca}). Center panel: profile of the layer-averaged probability of belonging to the HDA phase in the $M_2(O)$ model. Bottom panel: profile of the layer-averaged probability of belonging to the HDA phase in the $M_2(O,H)$ model.}
    \label{fig:interface}
\end{figure*}

The figure~\ref{fig:interface} shows the probability profile of belonging to the HDA phase $P(hda)$, averaged over a 2.0~\AA~thick layer. The number of molecules in each layer is not less than 100. If $P(hda)$  is close to 1, then this corresponds to the HDA structure ($100\leqslant X\leqslant 160$~\AA), if $P(hda)$ is close to zero, then to LDA ($0\leqslant X\leqslant 80$~\AA~and $180\leqslant X\leqslant 260$~\AA). In regions of approximately 80-100 and 160-180~\AA, the average probability has an intermediate value, which corresponds to the LDA~||~HDA interface. 

The interface thickness calculated via the Gibbs method (integral of the probability density) is approximately 11~\AA~for both models $M_2(O)$ and $M_2(O,H)$. Within this interfacial layer, about 20\% of the molecules exhibit prediction probabilities below 0.9, corresponding to intermediate amorphous ice. The rearrangement from pure LDA  to pure HDA thus spans approximately 3-4 coordination spheres. This indicates that the boundary is not atomically sharp (which would correspond to 1-2 molecular layers), nor is it macroscopically diffuse (which would extend over tens of angstroms).

\begin{figure*}[ht]
    \begin{minipage}[h]{0.49\linewidth}
    \centering
    \includegraphics[trim=1 1 1 1, clip, width=\textwidth]{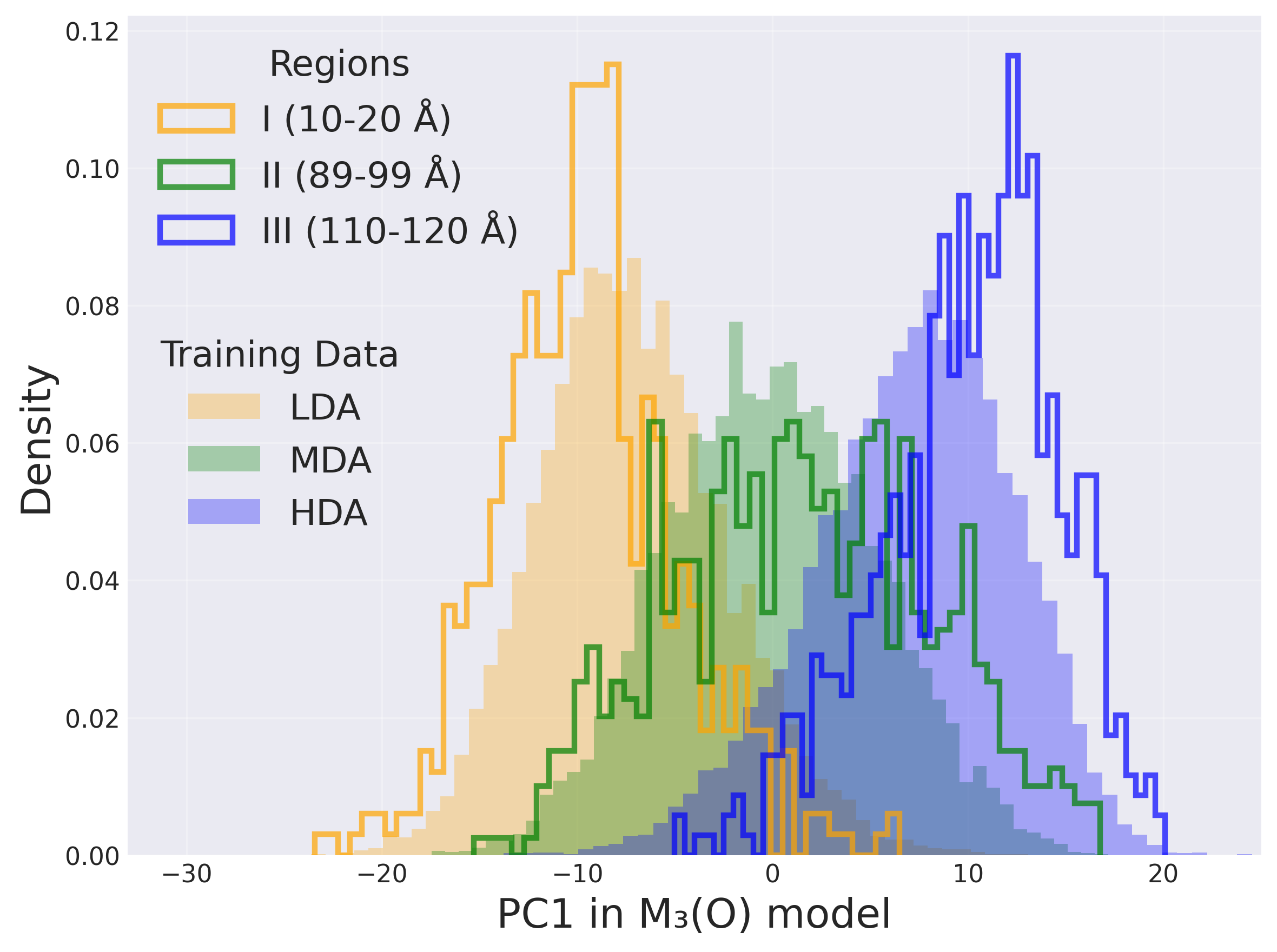} a)
    \end{minipage}
    \hfill
    \begin{minipage}[h]{0.49\linewidth}
    \centering
    \includegraphics[trim=1 1 1 1, clip, width=0.98\linewidth]{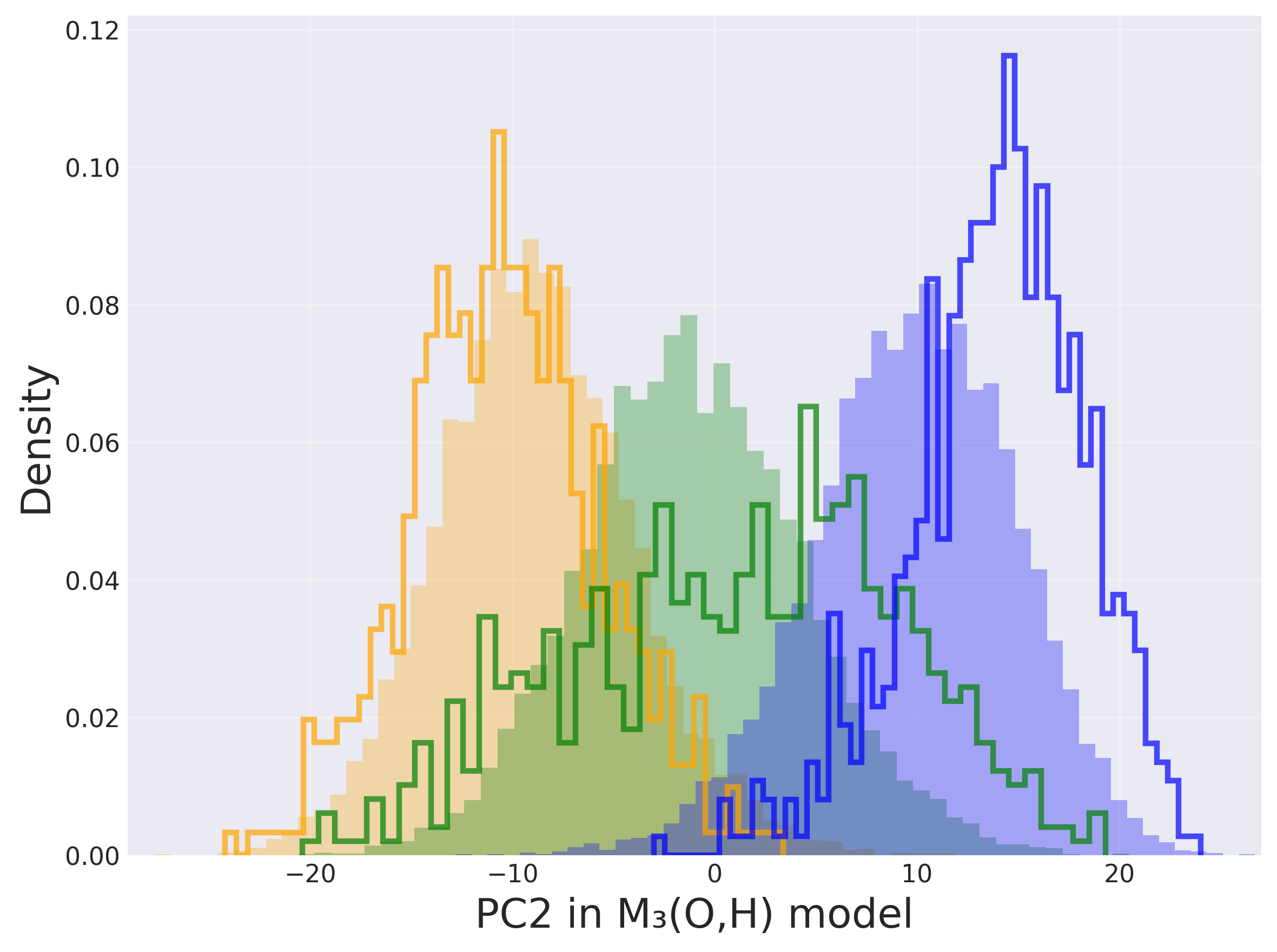} b)
    \end{minipage}
\caption{Distributions of the projected atomic environments in PC space. Training data (LDA, MDA, HDA) are shown together with atomic environments taken from three 10~\AA~thick layers of the LDA~||~HDA~||~LDA~ system (see Fig.~\ref{fig:interface}). a)~PC1 component of the M$_3$(O) model; b)~PC2 component of the M$_3$(O,H) model.}
\label{fig:interface_pca}
\end{figure*} 

Interestingly, the thickness defined by the distance between the 5\% and 95\% probability levels increases from approximately 11~\AA~in the $M_2(O)$ model to 18~\AA~in the $M_2(O,H)$. This discrepancy arises because the $M_2(O,H)$ model exhibits long tails in the probability profile: a small but non-zero probability of finding HDA-like environments persists in the LDA domain, and vice versa. The Gibbs method, being an integral measure, is less sensitive to these tails, whereas the 5–95\% distance directly captures their spatial extent. In particular, this broadening is not observed in the $M_2(O)$ model, where both measures produce approximately the same thickness. The presence of long tails in the $M_2(O,H)$ model suggests that the hydrogen-bonding network senses the approaching phase boundary over a longer range than the oxygen sublattice. In other words, the orientational order begins to change well before the density changes, indicating a two-stage relaxation process across the interface. This interpretation is consistent with the slow convergence of the interface profile (see Supplementary Fig.~S5) and with the notion that hydrogen-bond reorganization is kinetically hindered at low temperatures. 


To directly test whether interfacial molecules possess structural characteristics similar to bulk MDA, we project the atomic environments from three distinct regions of the system onto the principal component space obtained from the training set of the three-class classifiers $M_3(O)$ and $M_3(O,H)$, which includes LDA, HDA, and MDA structures (see Section~\ref{sec:model_learning}). For the $M_3(O)$ model, we examine the distributions along the PC1 component, and for $M_3(O,H)$, along the PC2 component. This choice is driven by data separability (see Fig.~\ref{fig:pca}). The first layer (X~=~10-20~\AA) is pure LDA, the second layer (X~=~89-99~\AA) is the interface, and the third layer (X~=~110-120~\AA) is pure HDA. 
As shown in Fig.~\ref{fig:interface_pca}, the atomic configurations from the first layer (the LDA slab) coincide with the LDA cluster in the PCA space, while those from the third layer (the HDA slab) coincide with the HDA cluster. 

The interfacial layer projects onto the same region as bulk MDA, albeit with a broader distribution and lower peak density. This broadening is expected, since the interfacial layer contains a mixture of MDA-like configurations and molecules in the process of transitioning between phases. Importantly, the PCA projection is computed solely from the training data without any information about the interface, yet it unambiguously places the interfacial environments in the MDA region of the descriptor space for both the $M_3(O)$ and $M_3(O,H)$ models. This result provides strong independent evidence that the interfacial layer is structurally similar to MDA (see also Supplementary Fig.~S6 online). 

\subsubsection{\label{sec:interface_response} Response of the LDA~||~HDA system to external influences}

\begin{figure}[ht]
    \begin{minipage}[h]{0.99\linewidth}
    \centering
        \includegraphics[trim=1 30 1 1, clip, width=\textwidth]{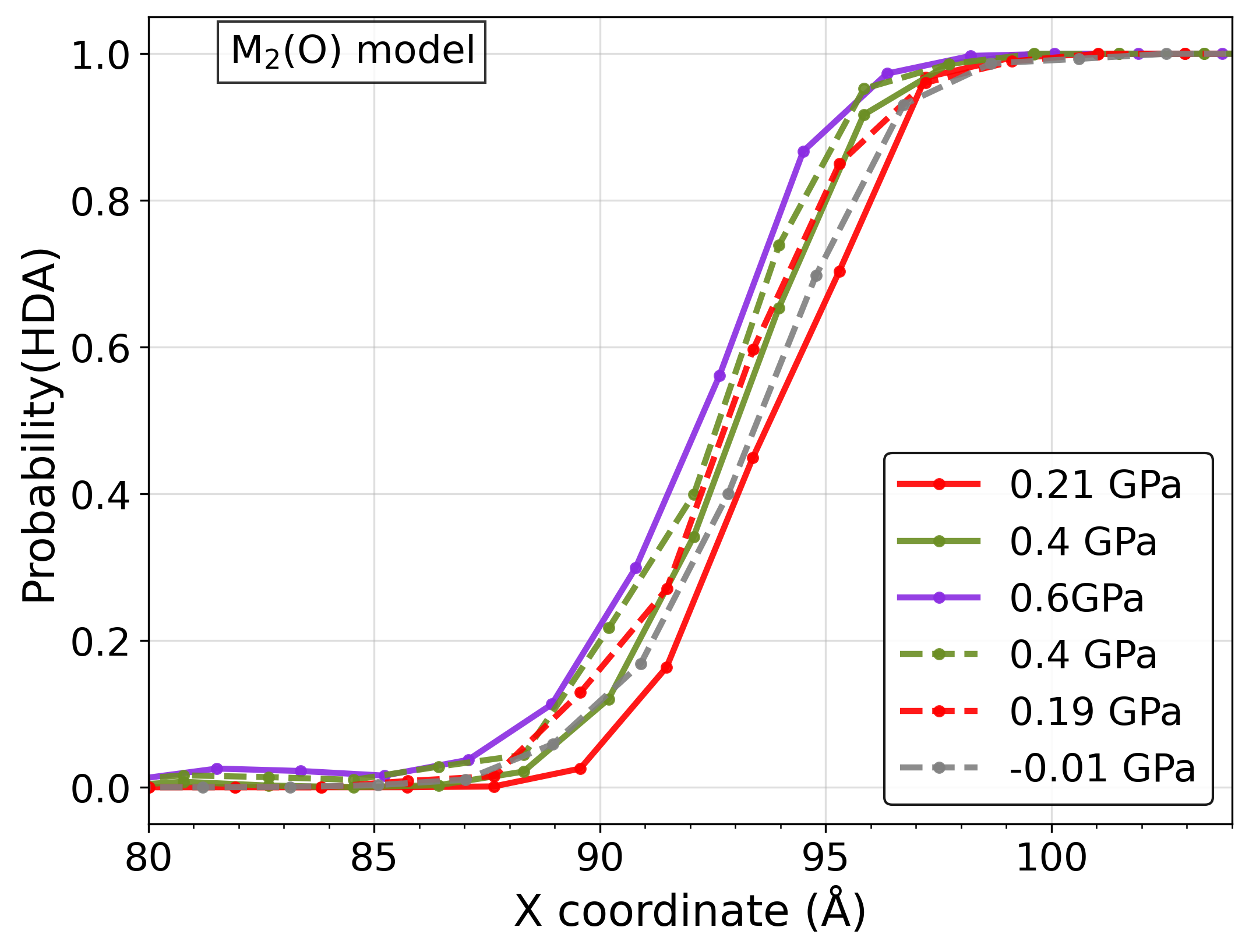}
    \end{minipage}
    \begin{minipage}[h]{0.99\linewidth}
    \centering
        \includegraphics[trim=1 1 1 1, clip, width=\linewidth]{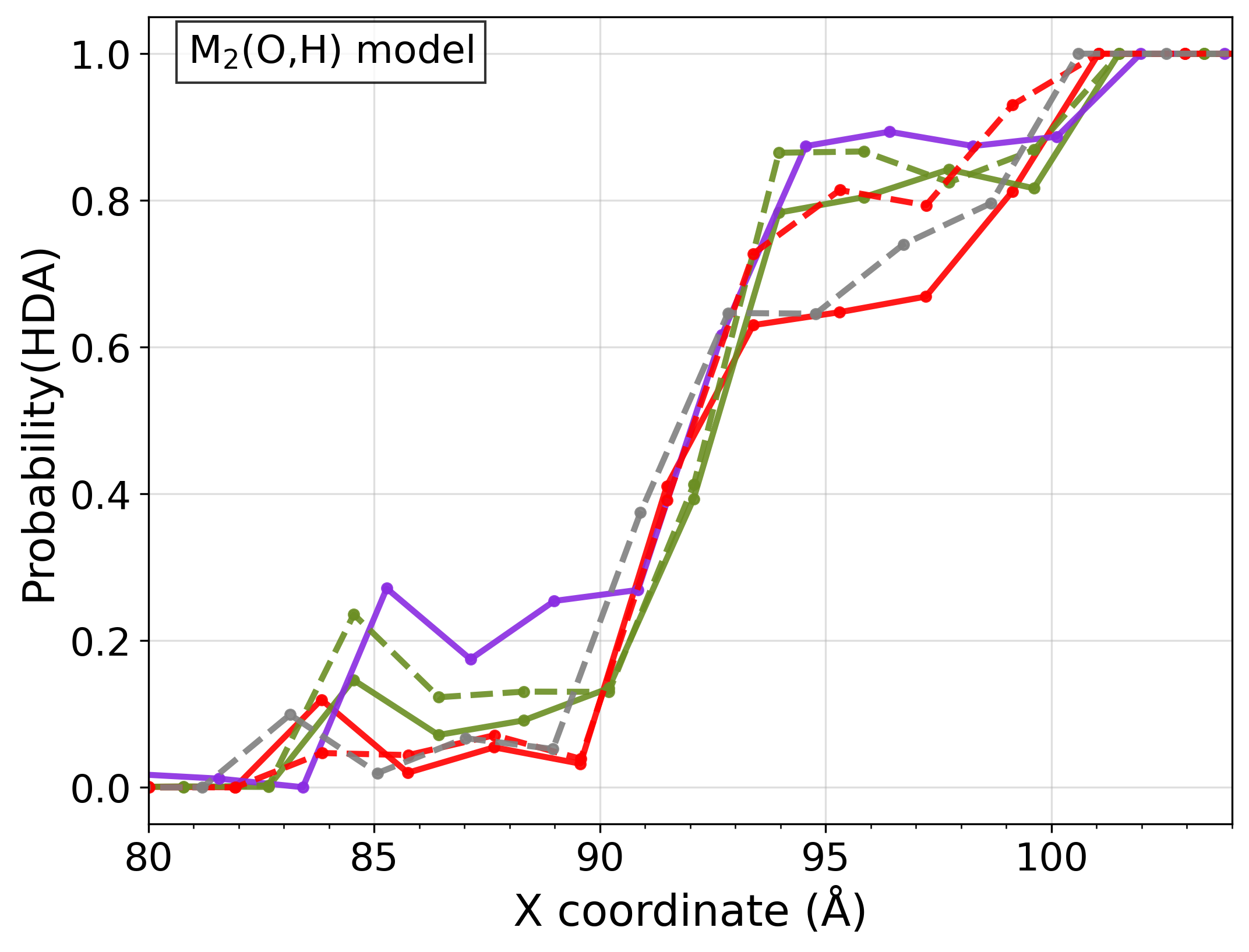}
    \end{minipage}
\caption{Probability profile at different stages of compression up to 0.6~GPa along the axis perpendicular to the phase boundary (solid lines) and at decompression stages (dashed lines). The result for model $M_2(O)$ is shown in the upper panel, for $M_2(O,H)$ -- in the lower panel.}\label{fig:interface_response}
\end{figure}

To initiate interface blurring and observe the transition to intermediate amorphous states, an external force must be applied to the system that disturbs it from the equilibrium (or metastable) state. There are three main levers: temperature, pressure, and shear. Each of these will yield a different physical scenario. 

In this paper, we consider the response of the LDA~||~HDA interface to compression and decompression along the X-axis, perpendicular to the phase boundary plane. The compression rate is 0.01~GPa/ns. Figure~\ref{fig:interface_response} shows how the profile changes with increasing pressure. Two observations are particularly noteworthy. First, the interface thickness remains virtually constant even under significant deviation from the initial pressure. Second, the relative fractions of LDA, HDA, and IA molecules show no substantial redistribution. However, the interface shifts toward the LDA phase by approximately 1-2~\AA, consistent with the higher stability of HDA in this pressure range. During decompression back to 0.4~GPa, the interface returns to the position observed during compression at the same pressure. However, upon further decompression to 0.2~GPa, the interface does not fully recover its original position; a residual shift of approximately 1~\AA~remains. Full recovery is achieved only after decompression to 0~GPa, i.e., below the initial equilibrium pressure. This indicates a moderate kinetic hysteresis, suggesting that the interface requires an underpressure to fully relax after a compression cycle. 

\section{\label{sec:discussion}Discussion}
The results presented above provide a detailed microscopic picture of the LDA~$\to$~HDA transformation and the nature of the intermediate states involved. In this section, we discuss the implications of our findings for understanding polyamorphic transitions in water, compare them with existing experimental and computational studies, and outline the limitations and perspectives of our approach.

\subsection{The features of the local structure of amorphous ices}

The use of SOAP descriptors provides a systematic and unbiased view of the local environment. The feature importance analysis and PCA decomposition presented in Section~\ref{sec:model_learning} provide insight into the structural characteristics that distinguish LDA, HDA, and MDA.

In the oxygen-only $M_3(O)$ model, classification relies mainly on local density and tetrahedral distortion. Including hydrogen atoms shifts discriminative information from PC1 to PC2, demonstrating that hydrogen atoms carry essential orientational information that is not captured by oxygen positions alone.

The redundancy of oxygen-oxygen correlations becomes evident when we train a classifier using only O-H and H-H features (i.e., excluding O-O). This hydrogen-bond-only model achieves classification performance nearly identical to that of the full $M_3(O, H)$ model (see Supplementary Table~SI). This confirms that orientational information encoded in the hydrogen network is the primary driver of phase discrimination, whereas oxygen packing plays a secondary role. The importance of hydrogen bond orientation as a distinct structural descriptor is further supported by recent work showing that the thermal expansivity of intermolecular hydrogen bonds in water differs significantly from that of the average O--O distance, demonstrating that orientational degrees of freedom carry information not captured by the oxygen sublattice alone~\cite{gabriel2024local}. 

The permutation importance ranking in Table~\ref{tab:pairwise_top7} shows that different pairs of phases are distinguished by different structural features:

\textbf{HDA~vs.~LDA.} The key distinguishing features are OH-type features with high angular moments (OH\_256, OH\_254, OH\_255), probing hydrogen-bond orientation correlations between the second and third coordination shells (at distances of approximately 4-6~\AA). This aligns with the established view that LDA preserves a more tetrahedrally oriented hydrogen-bond network, while HDA exhibits disrupted orientational correlations due to interstitial molecules~\cite{szekely2016tetrahedrality, martovnak2005evolution, subbotin2006modelling, tong2017microstructural, montes2020structural}. 
The difference with the work of Gallagher et al.~\cite{gallagher2026local}, who identified local hydrogen density as a key descriptor for distinguishing LDA from HDA, likely reflects SOAP's greater sensitivity to angular information. 

\textbf{LDA~vs.~MDA.} Discrimination is dominated by HH\_000, indicating that the primary difference lies in local hydrogen packing. Differences in the orientational order of the oxygen sublattice (OO\_355, OO\_225) play a secondary but non-negligible role. The difficulty in distinguishing MDA from LDA is consistent with the findings of Faure Beaulieu et al.~\cite{faure2024high}, who reported a 18-23\% misclassification rate between these two phases using a BOO-based neural network.

\textbf{HDA~vs.~MDA.} The top features are OO\_226 and OH\_226, followed by OO\_136 and OH\_135. This indicates that high angular modes in both oxygen-oxygen and oxygen-hydrogen correlations dominate, pointing to differences in orientational order within the second coordination shell.

Taken together, the feature importance rankings reveal an asymmetric relationship between the three phases. The structural difference between LDA and MDA is dominated by hydrogen packing, whereas the difference between HDA and MDA is governed by oxygen medium-range order (OO features of the second shell). This asymmetry suggests that MDA is not simply a compressed form of LDA or a diluted form of HDA. Rather, MDA possesses its own structural identity, with hydrogen packing distinguishing it from LDA and oxygen medium-range order distinguishing it from HDA.

The intermediate position of MDA in PCA space, combined with the asymmetric feature rankings, supports the continuum interpretation~\cite{eltareb2024continuum}, where intermediate-density amorphous ices occupy regions of the potential energy landscape between LDA and HDA basins. Crucially, the feature importance ranking for HDA-LDA discrimination remains unchanged between three-class and binary models, confirming that these features are intrinsic and not artifacts of MDA in the training set. This justifies using the binary classifier for analyzing the compression trajectory.

\subsection{The nature of intermediate amorphous states: interfacial layer versus bulk phase}

One of the central questions addressed in this work concerns the structural role of intermediate amorphous states in the LDA$\to$HDA transition~\cite{ramesh2024microscopic, faure2024high}. 
In this specific context, we found that these amorphous configurations do not form a bulk intermediate phase, but are localized at the interface between the HDA and LDA domains.

The results obtained in this study are consistent with the continuum concept proposed by Eltareb et al.~\cite{eltareb2024continuum} and add a spatial dimension to it: in a phase-separated system undergoing a first-order phase transition, a continuum of structures is realized spatially. When moving across the interface from the LDA side to the HDA side, the local structure continuously evolves from LDA-like to HDA-like, passing through intermediate configurations. Moreover, MDA is one representative of a family of intermediate amorphous states.

Several lines of evidence support this conclusion. 
\begin{itemize}
 \item Molecules classified as IA (those with confidence below the threshold) are not randomly distributed, but form a spatially connected layer between LDA and HDA domains.  
 \item The proportion of such molecules reaches a maximum precisely at the midpoint of the transition, where the total interfacial area between the two phases is greatest. 
 \item As shown by the planar interface analysis, the thickness of this interfacial layer remains constant (10-12~\AA, corresponding to 3-4 molecular layers) over a wide pressure range, suggesting that it represents an equilibrium property of the LDA~||~HDA interface rather than a kinetically trapped transient state. 
 \item We directly compare the local environments of the interfacial layer with those of bulk MDA using principal component analysis (Fig.~\ref{fig:interface_pca}). The projected points from the pure LDA and HDA slabs coincide with their respective reference clusters, while the interfacial layer projects onto the same region as bulk MDA. Since the PCA projection is computed solely from the training data without any interface information, this provides strong independent evidence that the local structure of the interfacial layer closely resembles that of bulk MDA.
\end{itemize}

Interestingly, the  $M_3(O)$ and $M_3(O, H)$ models yield qualitatively similar interface patterns: both localize IA molecules at the phase boundary and give consistent Gibbs thickness values. However, the $M_3(O, H)$ model reveals long tails in the probability profile that are absent in the oxygen-only model (Fig.~\ref{fig:interface}). This indicates that the hydrogen-bond network undergoes a more broad and gradual transition than the oxygen sublattice. This observation aligns with the work of Gabriel et al.~\cite{gabriel2024local}, who demonstrated that the hydrogen-bond network in water encodes structural information beyond the oxygen sublattice — a property that persists from the liquid to the glassy state. The fact that both models, despite their differences, lead to the same conclusion about the interfacial nature of IA states confirms the robustness of our results.

It should be noted that Gallagher et al.~\cite{gallagher2026local}, using a probabilistic classifier based on ACSF and BOO descriptors, found no evidence for intermediate structures in the LDA~$\to$~HDA transition.
The most likely explanation for this discrepancy lies in the different sensitivities of the structural descriptors. While the selected ACSF and BOO features are informative, the SOAP descriptor provides a more complete and unbiased representation of the local atomic environment. This may allow it to capture subtle structural signatures of the interfacial layer that are partially averaged out in the lower-dimensional feature space. This comparison highlights the importance of descriptor selection in machine learning-based structural analysis of amorphous systems. 
Moreover, their simulations used potentials obtained through machine learning (DP-SCAN and DP-MBpol~\cite{szukalo2025computational}) rather than classical force fields. It is possible that the different potential energy surfaces accessed by these models influence the structural evolution of the interface, potentially suppressing or altering the formation of intermediate states. Thus, the observation of MDA-like intermediate states may depend on the specific water model used, which warrants further systematic investigation across different force fields.

\subsection{Mechanism of LDA~$\to$~HDA transformation}
The model used in this work allows us to identify the phenomena of formation and growth of HDA domains, which confirms previous results of MD modelling~\cite{garkul2022molecular, dhabal2023kinetics, ramesh2024microscopic} and experimental observations~\cite{tonauer2017high}. The size of the critical nuclei before the rapid growth stage is 5–10 molecules. A sharp increase in cluster size occurs at approximately the same pressure regardless of the threshold value, confirming the robustness of the transition and the absence of an artifact of the classification procedure. Unlike our previous analysis, which relied on a simpler local density-based classification criterion and did not take into account intermediate molecules, the present approach provides a more accurate picture by explicitly identifying molecules with low classification confidence at the phase boundary. Nevertheless, the cluster growth dynamics obtained using the two methods are in good agreement, as shown in Fig.~\ref{fig:hda_cluster_evolution}. This confirms that the presence of an interphase layer does not fundamentally alter the mechanism of HDA phase nucleation and growth.

The observed spatial correlation between IA molecules and the phase boundary allows us to propose a detailed microscopic mechanism for the pressure-induced transformation of LDA~$\to$~HDA. The process occurs in three distinct stages:

(i)~At low pressures ($P\lesssim 0.5$~GPa), the system is predominantly in the LDA state, with rare small HDA clusters forming heterogeneously. The interfacial layer surrounding these nascent clusters consists of IA-like molecules whose local structure occupies a position intermediate between the two bulk phases. 

(ii)~As the pressure increases to the range of 0.6-0.7~GPa, the HDA clusters grow, engulfing the surrounding LDA matrix. At this stage, the total interfacial area -- and, consequently, the number of IA molecules -- reaches a maximum. Importantly, the thickness of the interfacial layer remains constant, indicating that growth occurs through interfacial advancement rather than expansion. 

(iii)~At high pressures ($P\gtrsim 0.8$~GPa), the system becomes predominantly HDA, with only isolated LDA residues and, correspondingly, a small population of molecules at the interfacial region.

This mechanism is consistent with the classical picture of a first-order phase transition proceeding through nucleation and growth, but with an important refinement: the interface between the coexisting phases is not atomically sharp, but has a finite thickness of approximately 10-12~\AA, within which the local structure differs from both bulk phases. 

The growth mechanism we observe, involving the formation of an interfacial layer before rapid HDA cluster coalescence, suggests a multistep nucleation pathway. This behavior is consistent with the independent findings of Ramesh, Mondal and Singh~\cite{ramesh2024microscopic}, who identified a pre-ordered intermediate phase during the LDA$\to$HDA transition in TIP4P/2005 and ST2 water models. In our case, IA molecules at the LDA-HDA boundary act as such precursors, initially appearing as a thin interfacial layer that then facilitates the subsequent growth and coalescence of HDA clusters. Phenomenologically similar multistep nucleation has also been reported in colloidal systems~\cite{takahashi2021multistep}, suggesting that such pathways may be more general than previously appreciated.


\subsection{Elastic response of the LDA~||~HDA interface}

An unexpected finding of this work is the systematic shift of the interface position under compression and decompression. When the pressure is increased from 0.2 to 0.6~GPa, the interface shifts by approximately 1-2~\AA~toward the LDA side, corresponding to the conversion of 1-2 molecular layers from LDA to HDA. Upon decompression back to 0.4~GPa, the interface returns to the position observed during compression at the same pressure. However, upon further decompression to 0.2~GPa, the interface does not fully recover its original position; a residual shift of approximately 1~\AA~remains. Full recovery is achieved only after decompression to 0~GPa, i.e., below the initial equilibrium pressure. This behavior indicates a moderate kinetic hysteresis: the interface requires an underpressure to fully relax after a compression cycle.

This observation has two important implications. First, the interface position is primarily determined by the thermodynamic conditions, but the system exhibits a memory effect, resulting in a slight hysteresis. Second, the interface thickness remains constant ($\approx 11$~\AA) throughout the entire compression-decompression cycle, independent of pressure and direction of change, suggesting that the interfacial structure itself is robust and does not broaden under external stress.

Similar interface migration phenomena have been observed in other amorphous systems under external stimuli. For instance, Moras et al.~\cite{moras2018shear, reichenbach2021solid} reported shear-induced amorphization and interface migration in silicon and diamond, a phenomenon they termed ``triboepitaxy''. Likewise, radiation-induced interface migration has been documented in Ti-Cr alloys~\cite{allen1987cascade}. Our observations are consistent with these findings, showing that interface migration under external pressure can occur without measurable broadening of the interfacial region.

\subsection{Broader perspectives and applications}

The methodology developed here -- combining SOAP descriptors with neural-network classification -- provides a general framework for analyzing local structure in disordered materials. It can be extended to other polyamorphic systems, such as silicon, germanium, or molecular glasses, where similar interface-mediated transitions may occur~\cite{brazhkin2011atomistic, durandurdu2002first, ganesh2009liquid}. More generally, the approach is applicable whenever local structural descriptors can be computed from atomistic simulations and a reference set of configurations is available for training. This includes studies of glass--glass transitions, pressure-induced amorphisation, and the identification of structural heterogeneity in supercooled liquids.

The application of methods for recognizing the structures of amorphous ices also has broad prospects in various fields. In cryoelectron microscopy~\cite{adrian1984cryo, lyu2023electron, lyumkis2019challenges}, for instance, amorphous ice serves as a matrix that preserves the native structure of biomolecules; understanding the local structure and stability of the ice matrix is essential for interpreting high-resolution images and avoiding beam-induced artifacts. Furthermore, the ability of amorphous ice to immobilize other molecules opens up new opportunities in cryopreservation and materials transport~\cite{nagy2020vitrification, rienzi2017oocyte}.

\section{\label{sec:conclusions}Conclusions}

In this work, we have combined SOAP descriptors with neural-network classification to investigate the pressure-induced LDA~$\to$~HDA transformation. Our analysis provides a molecular level space resolution characterization of the transformation in amorphous ice and resolves the question of the nature of intermediate amorphous (IA) states.

We demonstrate that the transformation proceeds via nucleation and growth of HDA clusters within the LDA matrix, with the phase boundary characterized by an interfacial layer of IA molecules of thickness 3--4 molecular layers. This interfacial layer is not a bulk phase, but rather a structurally distinct region that continuously connects the two amorphous states. Moreover, we show that this interfacial layer is structurally similar to MDA.

The LDA~||~HDA interface thickness remains constant under compression, while its position shifts reversibly. However, upon decompression, moderate kinetic hysteresis is observed: the interface partially recovers its position but requires decompression to lower pressures.

Feature importance analysis reveals that different phase pairs are distinguished by different structural characteristics:
\begin{enumerate}
\item HDA and LDA are primarily discriminated by OH angular descriptors probing hydrogen-bond orientations in the medium-range order;
\item LDA and MDA are distinguished by local hydrogen packing; 
\item HDA and MDA differ in the medium-range order of the oxygen sublattice. 
\end{enumerate}
These asymmetric fingerprints suggest that MDA possesses its own structural identity, rather than being simply compressed LDA or diluted HDA. Notably, the classifier trained only on hydrogen-related features (O-H and H-H) achieves the same accuracy as the full model (including O-O features), confirming that hydrogen bonds themselves reflect significant structural differences between the amorphous ices. 

\begin{acknowledgments}
We are grateful to Ekaterina Bagantsova and Prokhor Iashin for their help at the initial stage of this study, particularly for their contributions to developing and optimizing the neural network model. 
This work was prepared with the support of the Ministry of Science and Higher Education of the Russian Federation (State Assignment No.~075-00270-24-00). The authors gratefully acknowledge the access to the resources of the Supercomputer Center of JIHT RAS.
\end{acknowledgments}

\section*{Author Declarations}
\subsection*{Conflict of Interest}
The authors have no conflicts to disclose.

\subsection*{Author Contributions}
V.S. conceived the study, A.S. made the necessary calculations, processed the data and wrote the manuscript. Both authors analyzed the results and reviewed and edited the manuscript.

\section*{Data Availability Statement}

The data available from the authors upon a reasonable request.



\providecommand{\noopsort}[1]{}\providecommand{\singleletter}[1]{#1}%

\end{document}